\documentclass[12pt]{article} \def\theequation{\arabic{section}.\arabic{equation}}
\usepackage[nosort]{cite}
\newcounter{rown}

\usepackage[a4paper,hmargin=2cm,tmargin=5cm,bmargin=3cm]{geometry}
\usepackage{mathrsfs,eucal,graphicx}
\usepackage[centertags]{amsmath}
\usepackage{color}
\usepackage{amsfonts}
\usepackage{amssymb}
\usepackage{amsthm}
\usepackage{newlfont}
\begin{document}
\begin{flushright}
FTUV-09-0517 $\;$ IFIC/09-18\\
May 17, 2009
\end{flushright}
\vskip 2cm
\renewcommand{\theequation}{\arabic{section}.\arabic{equation}}
\begin{center}
{\Large\bf Cohomology of Filippov algebras and an \\analogue of Whitehead's lemma}

\vskip 2cm
 {J.A. de Azc\'{a}rraga, \\
{\it Dpto. de F\'{\i}sica Te\'orica and IFIC (CSIC-UVEG), University of Valencia, \\
 46100-Burjassot (Valencia), Spain}\\
 \vskip .5cm
 J. M Izquierdo,\\
{\it Departamento de F\'{\i}sica Te\'orica, Universidad de Valladolid, \\
47011-Valladolid, Spain}}
\end{center}
\vskip 2cm
\begin{abstract}
We show that two cohomological properties of semisimple Lie
algebras also hold for Filippov ($n$-Lie) algebras, namely, that
semisimple $n$-Lie algebras do not admit non-trivial central
extensions and that they are rigid {\it i.e.}, cannot be deformed
in Gerstenhaber sense. This result is the analogue
of Whitehead's Lemma for Filippov algebras.
A few comments about the $n$-Leibniz algebras case are
made at the end.
\end{abstract}
\newpage

\section{Introduction and outlook}

In the last years there has been an increasing interest in the
applications of various generalizations of the ordinary Lie algebra
structure to theoretical physics problems. In these generalizations,
which we shall denote generically as $n$-ary algebras, the two
entries of the standard Lie bracket are replaced by $n>2$ entries.
There are two main ways of achieving this,
depending on how the Jacobi identity (JI) of the ordinary Lie
algebras is looked at. The JI can be viewed as the statement that (a)
a double Lie bracket gives zero when antisymmetrized with respect to
its three entries or that (b) the Lie bracket is a
derivation of itself. Both (a) and (b) are equivalent for ordinary
Lie algebras and (a) is indeed an identity that follows from the
associativity of the composition of the Lie algebra generators.
\medskip

When a $n$-ary algebra is defined using the {\it characteristic
identity} that extends property (a) to a $n$-ary bracket, one
is led to a generalization denoted {\it higher order Lie algebras}
or {\it generalized Lie algebras} (GLA) $\mathcal{G}$
\cite{AzPePB:96b,AzBu:96}, and the characteristic identity satisfied by its
multibracket is called {\it generalized Jacobi identity} (GJI).
This generalization is natural for $n$ even (for $n$
odd, the $r.h.s.$ of the GJI, rather than being zero, is a larger
bracket with ($2n-1$) entries \cite{RACSAM:98}). Similar algebras have also been
discussed in \cite{Han-Wac:95,JLL:95,Gne:95,Mic-Vin:96, Gne:97};
GLAs may also be considered as a particular case (when
there is no violation of the GJI \cite{RACSAM:98}) of the
strongly homotopy algebras of Stasheff \cite{Lad.Sta:93,Lad.Mar:95, Sta:97, Be-La:09}.
When possibility (b) is used as the guiding principle, then one is led to
the {\it Filippov identity} (FI) \cite{Filippov}
and correspondingly to {$n$-Lie} or {\it Filippov
algebras} $\mathfrak{G}$ \cite{Filippov} (both terms, Filippov and $n$-Lie,
will be used indistinctly), for which the characteristic
identity is the FI. For $n=2$, both algebra structures
coincide and determine ordinary Lie algebras $\mathfrak{g}$;
when $n\geq 3$, the GJI ($n$ even) and the FI become
different characteristic identities and define, respectively,
generalized Lie algebras\footnote{The GLA were called {\it Lie $n$-algebras}
in \cite{Han-Wac:95}, where they were independently considered
(see also \cite{JLL:95}). But, rather than betting the
distinction between $n$-Lie algebras ($\equiv$ FA) and
and GLAs on the precise location
of a single letter ($n$-Lie alg. {\it vs.} Lie $n$-alg.),
we prefer our {\it higher order Lie algebras} or GLA
terminology.} (GLA) $\mathcal{G}$ and $n$-Lie or
Filippov algebras (FA) $\mathfrak{G}$.
\medskip

Filippov algebras \cite{Filippov, Kas:87, Kas:95a, Ling:93}
 have recently been found useful in the search for an
effective action describing the low energy dynamics of
coincident M2-branes or, more specifically,
in the Bagger-Lambert-Gustavsson model (BLG)
\cite{Ba-La:06,Ba-La:07,Gustav:08,Raam:08,Ba-La:08,Che-Sae:08,Go-Mi-Ru:08}.
The field theory BLG model contains scalar and fermion fields
that take values in a 3-Lie algebra plus gauge fields that are
valued in the adjoint representation of the Lie algebra
of the automorphisms of the 3-Lie algebra;
the model has $\mathcal{N}$=8 supersymmetries. The uniqueness of
the euclidean 3-Lie algebra $A_4$ was
in fact found in the context of the BLG model, where it
follows \cite{Ga-Gu:08, Pap:08} by assuming that the metric
needed for the BLG action has to be positive definite,
a condition that may be relaxed \cite{Go-Mi-Ru:08}
(see also \cite{deMe-Fi-M-E-Rit:09} and references
therein). We shall not discuss the BLG and related models
here; we shall just mention that
the original BGL action was subsequently reformulated
\cite{Raam:08} without using a three-Lie algebra,
and that other models for low energy multiple M2 brane dynamics
have appeared (albeit with $\mathcal{N}$=6 rather
than $\mathcal{N}$=8 supersymmetries) that do not use
a FA structure \cite{A-B-J-M:08} (see footnote 7). This paper
will be devoted instead to some purely mathematical
aspects of Filippov algebras. Other specific ternary
structures, such as Jordan-Okubo triple systems and
others will not be discussed here
(see \cite{Jac:49,Oku-Kam:96,Oku:03,Ker:08}
and references therein).
\medskip

It is well known that semisimple Lie algebras neither admit
non-trivial central extensions (see {\it e.g.} \cite{CUP}) nor
infinitesimal deformations \cite{Gers:63,Nij-Rich:67} so that they
are rigid or stable. This is so because their
cohomology groups $H_0^2(\mathfrak{g})$ and $H_\rho^2(\mathfrak{g})$
are trivial as a result of Whitehead's Lemma (see \cite{Jac:79}; in
fact, the $p$-th cohomology groups $H^p_\rho(\mathfrak{g},V)$, where
$V$ is a $\rho(\mathfrak{g})$-module, are zero for $\rho$
non-trivial\footnote{This is no longer true for $\rho=0$; for a simple
compact $\mathfrak{g}$, for instance, the fully antisymmetric
structure constants of the Lie algebra always determine a non-trivial
three-cocyle.} and $p\geq 0$). The standard proof uses that the
Cartan-Killing metric for semisimple algebras can be inverted and
then the inverse allows one to show that all $p$-cocycles are
$p$-coboundaries. We prove in this paper that both the triviality of
central extensions and the stability of semisimple Lie algebras
under deformations also hold for semisimple Filippov algebras, so that
these properties hold true for all $n\geq2$. We shall
show this by using a route that does not require introducing a generalized
Cartan-Killing form for the FA. The reason is twofold. First, the
Cartan-Killing bilinear form of a Lie algebra $\mathfrak{g}$ is
defined on its vector space. In contrast, its $n$-Lie algebra
analogue \cite{Kas:95a} (see eq.~\eqref{KasCK}), which one might
think of using as the Lie algebra Cartan metric to mimic
the proof there, is not a bilinear form on $\mathfrak{G}$ but on
their fundamental objects\footnote{There is still a parallel,
however, if one realizes that in both the $n=2$ and the arbitrary
$n$ cases the Killing metric is a trace form, namely
Tr$(ad_\mathscr{X}ad_\mathscr{Y})$. This relates it to the Lie
algebra of inner derivations of $n$-Lie algebras ($n\geq 2$) in
general.} $\mathscr{X}\in\wedge^{n-1} \mathfrak{G}$ (Sec.~2 below).
However, the semisimplicity criterion for a $n$-Lie algebra, which
states that the $2(n-1)$-linear Killing form generalization $k$
\cite{Kas:95a} in \eqref{KasCK} is not degenerate
(eq.~\eqref{intro7} below) does not guarantee
that $k$ is non-degenerate as a bilinear
form on $\wedge^{n-1} \mathfrak{G}$. Secondly, all simple
$n>2$ Filippov algebras are known \cite{Ling:93, Filippov} and they
are few when compared with the plethora of the
$n=2$ Cartan cassification of simple Lie algebras
and, further, all of them have the same general structure.
Specifically, the only simple real Filippov algebras
are the $(n+1)$-dimensional $n$-Lie algebras of type $A_{n+1}$
(eq.~\eqref{simple}) \cite{Ling:93, Filippov}, which may be thought
of as $n>2$ generalizations of the $n=2$ $so(3)$ and $so(1,2)$
ordinary Lie algebras. We shall take advantage of this
fact to show first the triviality of the central
extensions and deformations of simple $n$-Lie algebras; then,
using that any semisimple Filippov algebra is the direct sum
of its simple ideals \cite{Ling:93}, we shall extend the result to
semisimple $n$-algebras as well.
\medskip

As already mentioned, the central extension and deformation problems
for Lie algebras are formulated in terms of the second Lie algebra
cohomology groups (see {\it e.g.} \cite{CUP}) for the trivial and
the adjoint action respectively. Non-trivial central
extensions are characterized by non-trivial two-cocycles in
the Lie algebra cohomology group
$H^2_0(\mathfrak{g})=Z^2_0(\mathfrak{g})/B^2_0(\mathfrak{g})$
for the trivial action, whereas non-trivial infinitesimal
deformations require $H_\rho^2\not=0$ for $\rho=ad$. In the Filippov
algebras case, the generalization is not immediate, and in fact we
will show that the cocycles responsible for both extensions and
deformations may be considered as one-cocycles rather than two-cocycles. The
characterization of the cochains and the corresponding cohomology
complexes will be given in Secs.~4.1 and 5.1. It will turn out that
the cohomology complex adapted to the deformation problem obtained
in Sec.~4.1 is essentially equivalent to the one introduced by
Gautheron \cite{Gau:96} (see also \cite{Da-Tak:97,Rot:05,Tak:94}),
who was the first to consider the full
deformation cohomology complex for Nambu algebras.
\medskip

  Nambu algebras are, in fact, a particular case of $n$-Lie
algebras. Their $n$-bracket is provided by the Jacobian
determinant of $n$ functions or Nambu bracket \cite{Nambu:73},
although Nambu did not write the characteristic identity satisfied
by his ($n=3$) bracket, which is none other than the FI. This was
done in \cite{Filippov,Sa-Va:92,Sa-Va:93, Tak:93,Fil:98},
and {\it Nambu-Poisson structures} (N-P) have been much studied since
Nambu's original paper \cite{Nambu:73} and Takhtajan general study
\cite{Tak:93}, see \cite{Mu-Sud:76, Sa-Va:93,Cha:95, Cha-Tak:95,
Ale.Guh:96, Hie:97, Da-Tak:97, Ma-Vi-Vi:97,Vai:99, Mi-Vi:00, Cu-Za:02}
(ref.~\cite{Da-Tak:97} also considers Nambu superalgebras).
In fact, since the earlier considerations of $p$-branes as
gauge theories of volume preserving diffeomorphisms
\cite{Be-Se-Ta-To:90}, the infinite dimensional FAs given
by Nambu brackets have reappeared in applications
to brane theory \cite{Ho:96} and, in particular, in
the Nambu three-bracket realization of the mentioned
BLG model as a gauge theory associated with volume
preserving diffeormorphisms in a three-dimensional
space; see, in particular,
\cite{Ho-Hou-Ma:08, Sochi:08, Ho-Ma:08, Ba-To:08}.
\medskip

Much in the same way the Nambu-Poisson structures follow
the pattern of FAs, it is also possible to introduce {\it generalized
Poisson structures} (GPS) \cite{AzPePB:96a, AzPePB:96b, AIP-B:97}
(see further \cite{Mi-Vi:00,Iba.Leo.Mar.Die:97,Iba.Leo.Mar:97})
whose $n$-even {\it generalized Poisson brackets} (GPB)
satisfy the GJI and correspond to the  GLAs earlier
mentioned. This can also be achieved in the graded case,
which corresponds to graded GLAs \cite{A-I-Pe-PB:96}.
Note, however, that besides the two
properties that each Poisson generalization share respectively with
the GLAs and FAs (skewsymmetry of both $n$-ary Poisson brackets plus
the GJI (FI) for the GP (N-P) structures, respectively),
the $n$-ary brackets of both GPS and of N-PS satisfy an
additional condition, Leibniz's rule. There has been an extensive
discussion since the papers by Nambu \cite{Nambu:73} and
Takhtajan \cite{Tak:93} about the
difficulties of quantizing the N-P strucures. We shall not
touch the point of quantizing $n$-ary Poisson structures
here and will just refer instead to the papers above
and {\it e.g.}, to
\cite{Ho:96, AIP-B:97,Stern:98, Cu-Za:03b,Cu-Fa-Ji-Me-Za:09}
and references therein.
\medskip

The plan of this paper is as follows. Sec.~2 reviews some
facts on FAs in a (we hope) transparent notation. In Sec.~3
we show that Kasymov's analogue \cite{Kas:95a} of
the Cartan-Killing form for a FA $\mathfrak{G}$
(eq.~\eqref{KasCK} below), when viewed as a {\it bilinear} form on
$\wedge^{n-1}\mathfrak{G}$, is degenerate when $\mathfrak{G}$ is
semisimple. This is because the fundamental objects (Sec.~2)
$\mathscr{X}\in\wedge^{n-1}\mathfrak{G}$ may involve elements of
$\mathfrak{G}$ in different simple ideals (something that cannot
happen in the Lie algebra case, where the fundamental objects reduce
to single elements $X\in\mathfrak{g}$), although it is
non-degenerate when $\mathfrak{G}$ is simple. In Sec~4.1 we derive
the conditions for the existence and triviality of a central
extension of a Filippov algebra, which allows us to define
one-cocycle and one-coboundary conditions respectively. By
extending them to higher order cochains, this leads us naturally
to the expression of the corresponding cohomology complex,
which is given explicitly in that section. Sec.~4.2 contains the
proof of the triviality of all central extensions of semisimple
Filippov algebras. In Sec.~5.1 we derive the cohomological
conditions that govern the infinitesimal
deformations of FAs, and subsequently we obtain from
them the action of the coboundary operator and the cohomology
complex for the non-trivial action, which is the relevant one for
deformations of FAs;  both the left and the right
actions appear naturally in its definition.
Sec.~5.2 contains the proof of the rigidity of the
semisimple Filippov algebras. Sec.~6 is makes some observations
relating the FA and $n$-Leibniz algebra cohomologies
($n$-Leibniz algebras share the derivation property of
the FI with the FAs but not the total antisymmetry
of their $n$-brackets). Finally,
Sec.~7 presents some remarks concerning the extension
of the above results to $n$-Leibniz algebras.

\section{Filippov algebras: some basic definitions and properties}

We present in this section some salient features of FA in a form
that will be convenient for applications later.
\medskip

A FA algebra $\mathfrak{G}$ is a vector space endowed with a
skew-symmetric, $n$-linear bracket,
\begin{equation}
       (X_1,\dots , X_n)\in \mathfrak{G} \times\dots \times  \mathfrak{G}
       \mapsto [X_1,\dots , X_n] \in \mathfrak{G}
\label{intro1}
\end{equation}
that satisfies the FI,
\begin{equation}
\label{intro2}
       [X_1,\dots , X_{n-1},[Y_1,\dots Y_n]] = \sum_{a=1}^n
       [Y_1,\dots Y_{a-1}, [X_1,\dots , X_{n-1}, Y_a ], Y_{a+1}
       ,\dots Y_n] \ ,
\end{equation}
which states that the bracket $[X_1,\dots,X_{n-1},\;\,]$, where the
last entry is empty, is a derivation of the FA.

Given a basis $\{X_{a_i}\}$
of the $\mathfrak{G}$ vector space, the FA is
characterized by its structure constants,
\begin{equation*}
[X_{a_1},\dots , X_{a_n}] =  f_{a_1\dots a_n}^b \,X_b \; , \quad a_i=1,\dots,n \quad  .
\end{equation*}
\medskip

The properties of $n$-Lie algebras have been studied, along the
lines of Lie algebra theory, by Filippov \cite{Filippov, Fil:98}, Kasymov
\cite{Kas:87, Kas:95a}, Ling \cite{Ling:93} and others. For
instance, a subspace $I\subset \mathfrak{G}$ is an {\it ideal} of
$\mathfrak{G}$ if
\begin{equation*}
[X_1,\dots,X_{n-1},Z]  \subset I \quad  \forall X\in
\mathfrak{G}\,,\,\forall Z\in I \quad.
\end{equation*}
The above bracket may be rewritten in the form
\begin{equation}
\label{adZ}
 [X_1,\dots , X_{n-1},Z] := \mathscr{X}\cdot Z \equiv [\mathscr{X},Z]
 \equiv ad_{\mathscr{X}} Z \; ,
\end{equation}
where\footnote{The notation $\mathscr{X}\in \wedge^{n-1}\mathfrak{G}$
reflects that the fundamental object $\mathscr{X}=(X_1,\dots,X_{n-1})\in \mathfrak{G}\times
\mathop{\dots}\limits^{n-1}\times \mathfrak{G}$ is antisymmetric
in its arguments and does not imply that $\mathscr{X}$ is a
$(n-1)$-multivector obtained by the associative wedge product of
vector fields.} $\mathscr{X}\in\wedge^{n-1}\mathfrak{G}$. The
objects $\mathscr{X} \in \wedge^{n-1} \mathfrak{G}$ play an
important r\^ole in the theory of FA, and accordingly we shall call
them \textit{fundamental objects}. In fact, the properties of
FA are largely determined by them; they also determine derivations
that generate an associated Lie algebra. The fundamental objects are
characterized by ($n-1$) elements $(X_1,\dots,X_{n-1})$ of $\mathfrak{G}$ and are
skewsymmetric in them. In terms of these fundamental objects
$\mathscr{X}$, the FI may be rewritten as:
\begin{equation}
  \mathscr{X} \cdot [Y_1,\dots, Y_n] = \sum_{a=1}^n [Y_1,\dots ,
  \mathscr{X}\cdot Y_a, \dots , Y_n]\quad \textrm{or}\quad  ad_{\mathscr{X}}
  [Y_1,\dots, Y_n] = \sum_{a=1}^n [Y_1,\dots ,
  ad_{\mathscr{X}} Y_a, \dots , Y_n]\ .
\label{intro2a}
\end{equation}
Thus, the FI just reflects that
$ad_{\mathscr{X}}\equiv\mathscr{X}\cdot\,\equiv[X_1,\dots,X_{n-1},\quad]$
is a derivation of the FA (which may be called inner since it is
determined by elements of $\mathfrak{G}$). For the particular case
of an ordinary Lie algebra $\mathfrak{g}$, $n=2$, $\mathscr{X}=X$ and thus
$ad_{\mathscr{X}}\in \hbox{End}\,\mathfrak{G}$ reduces to the
standard adjoint derivative $ad_X\in\hbox{End}\,\mathfrak{g}$.
\medskip

A FA is {\it simple} if
$[\mathfrak{G},\dots,\mathfrak{G}]\not=\{0\}$ and has no ideals
different from the trivial ones, $\{0\}$ and $\mathfrak{G}$. A
$n$-Lie algebra is {\it semisimple} if it has no solvable ideals, an
ideal $I\subset \mathfrak{G}$ being {\it solvable} \cite{Filippov}
if the following sequence of ideals,
\begin{equation}
   I^{(0)}:=I\ ,\quad I^{(1)}:=[I^{(0)},\dots, I^{(0)}] \ ,\dots,\
   I^{(s)}:=[I^{(s-1)},\dots, I^{(s-1)}] \ , \dots
\label{intro4a}
\end{equation}
ends {\it i.e.}, there exists an $s$ for which
$I^{(s)}=0$\footnote{The solvability notion for Lie algebras allows
for various generalizations when moving to FAs, $n>2$, because the
$n$-bracket has more than two entries. For a $n$-Lie algebra the
notion of $k$-solvability was introduced by Kasymov \cite{Kas:87}
(see also \cite{Ling:93}) by taking
$\mathfrak{G}^{(0,k)}=\mathfrak{G}\;,\; \mathfrak{G}^{(m,k)}=
[\mathfrak{G}^{(m-1,k)},\dots,\mathfrak{G}^{(m-1,k)},\mathfrak{G},\dots,\mathfrak{G}]$,
where there are $k$ entries $\mathfrak{G}^{(m-1,k)}$ at the
beginning of the $n$-bracket. Filippov's solvability
\cite{Filippov}, used above, corresponds to $k$-solvability for
$k=n$; $k$-solvability is stronger and
implies $n$-solvability for all $k$ \cite{Ling:93}.}.
Kasymov's
generalization \cite{Kas:95a} of the Cartan criterion then states
that a $n$-Lie algebra is semisimple iff the following
($2n-2$)-linear generalization of the Cartan-Killing form
\begin{equation}
\label{KasCK} k(\mathscr{X},\mathscr{Y}) = k(X_1,\dots ,
X_{n-1},Y_1,\dots ,Y_{n-1}) :=Tr(ad_{\mathscr{X}}ad_{\mathscr{Y}})
\end{equation}
is non-degenerate, {\it i.e.} iff
\begin{equation}
\label{intro7}
        k(Z, \mathfrak{G},{\mathop{\dots}\limits^{n-2}},\mathfrak{G},
        \mathfrak{G}, {\mathop{\dots}\limits^{n-1}}, \mathfrak{G})=0
        \ \Rightarrow \ Z=0 \,
\end{equation}
where the $2n-3$ arguments besides $Z$ are arbitrary elements of
$\mathfrak{G}$.
\medskip

 It is convenient to introduce a composition law for
fundamental objects $\mathscr{X} = (X_1,\dots , X_{n-1})$,
$\mathscr{Y} = (Y_1,\dots , Y_{n-1})$,
$X_i,Y_i\in\mathfrak{G}\;,\; i=1,\dots,(n-1)$. The
{\it composition $\mathscr{X}\cdot \mathscr{Y}$} is given by the sum of
fundamental objects
\begin{eqnarray}
\label{intro4}
        \mathscr{X}\cdot \mathscr{Y} := \sum_{a=1}^{n-1}
        (Y_1,\dots ,Y_{a-1} ,[X_1,\dots , X_{n-1}, Y_a], Y_{a+1},
        \dots, Y_{n-1}) \; .
\end{eqnarray}
For a $n=2$ FA or ordinary Lie algebra, $\mathscr{X}\cdot \mathscr{Y}$ reduces
to $X\cdot Y=[X,Y]$. With the above notation, the following Lemma
follows from the FI:
\medskip

\noindent
{\bf Lemma} ({\it Properties of the composition of fundamental
objects})

 The dot product of fundamental objects
$\mathscr{X}$ of a $n$-Lie algebra $\mathfrak{G}$ satisfies the relation
\begin{equation}
\label{Lie-setb}
\mathscr{X}\cdot (\mathscr{Y}\cdot \mathscr{Z}) -
\mathscr{Y}\cdot (\mathscr{X} \cdot \mathscr{Z}) =
(\mathscr{X}\cdot\mathscr{Y}) \cdot \mathscr{Z} \qquad \forall
\mathscr{X}, \mathscr{Y}, \mathscr{Z} \in \wedge^{n-1}\mathfrak{G} \; .
\end{equation}
As a result, the images $ad_{\mathscr{X}}$ of the fundamental
objects by the the adjoint map $ad:\wedge^{n-1}\mathfrak{G}
\rightarrow \hbox{InDer}\,\mathfrak{G}$ determine (inner)
derivations of the FA that satisfy\footnote{Notice that $Z$ in
eq.~\eqref{Lie-n-Lie} may be replaced in general by $v\in V$,
$\mathscr{X}\cdot v:= \rho(\mathscr{X})\cdot v$, where $\rho$ is the
action that makes the vector space $V$ a
$\rho(\mathfrak{G})$-module.}
\begin{equation}
\label{Lie-n-Lie}
\begin{aligned}
\mathscr{X}\cdot(\mathscr{Y}\cdot Z)-& \mathscr{Y}\cdot(\mathscr{X}\cdot Z)
=(\mathscr{X}\cdot\mathscr{Y})\cdot Z \quad  \mathrm{or} \\
\qquad
 ad_{\mathscr{X}}ad_{\mathscr{Y}} Z - ad_{\mathscr{Y}}  & ad_{\mathscr{X}} Z =
ad_{\mathscr{X}\cdot\mathscr{Y}} Z  \qquad \forall
\mathscr{X}, \mathscr{Y} \in \wedge^{n-1}\mathfrak{G} \,,\,\forall \; Z\in\mathfrak{G} \quad .
\end{aligned}
\end{equation}

{\it Proof}:
To prove the assertion, it is sufficient to check
eq.~\eqref{Lie-setb}. Let us compute first $\mathscr{X}\cdot
(\mathscr{Y}\cdot \mathscr{Z})$. This is given by
\begin{equation*}
\begin{aligned}
 \mathscr{X}  &\cdot (\mathscr{Y}\cdot \mathscr{Z})=
\sum_{i=1}^{n-1} \mathscr{X} \cdot  \left(
Z_1,\dots,Z_{i-1},[Y_1,\dots,Y_{n-1},Z_i],Z_{i+1},\dots, Z_{i-1} \right) \\
 &= \sum_{i=1}^{n-1}  \sum_{j \not=i, \,j=1}^{n-1} \left(
 Z_1,\dots,Z_{j-1},[X_1,\dots,X_{n-1},Z_j],\ Z_{j+1},  \dots,
 Z_{i-1},\ [Y_1,\dots,Y_{n-1},Z_i],Z_{i+1},\dots,Z_{n-1}\right)\\
& \quad + \sum_{i=1}^{n-1} \left(
Z_1,\dots,Z_{i-1},[X_1,\dots,X_{n-1},[Y_1,\dots,Y_{n-1},Z_i]],
 Z_{i+1},\dots, Z_{n-1} \right) \; .
\end{aligned}
\end{equation*}
The first term in the $r.h.s$ is symmetric in $\mathscr{X},
\mathscr{Y}$; hence,
\begin{equation}
\begin{aligned}
\label{n-L-s} & \mathscr{X}\cdot (\mathscr{Y}\cdot \mathscr{Z}) -
\mathscr{Y}\cdot (\mathscr{X} \cdot \mathscr{Z})  =\\
\sum_{j=1}^{n-1} & \left( Z_1, \dots,Z_{j-1}, \{\,
[X_1,\dots,X_{n-1}, [Y_1,\dots,Y_{n-1},Z_j]] - [Y_1,\dots,
Y_{n-1},[X_1,\dots,X_{n-1},Z_j ]]\, \}, Z_{j+1},\dots,Z_{n-1}
\right) \; .
\end{aligned}
\end{equation}
On the other hand, using definition \eqref{intro4}, we find
\begin{equation}
(\mathscr{X} \cdot \mathscr{Y} )\cdot \mathscr{Z}=
\sum_{j=1}^{n-1}\sum_{i=1}^{n-1} (Z_1,\dots,Z_{j-1},
[Y_1,\dots,[X_1,\dots,X_{n-1},Y_i],\dots,Y_{n-1},Z_j],Z_{j+1},\dots,Z_{n-1}).
\end{equation}
Now, using the FI for $[X_1,\dots,X_{n-1},[Y_1,\dots,Y_{n-1},Z_j]]$,
we see that the above expression reproduces \eqref{n-L-s}.

 Obviously, the above proof carries forward for the simplest case where the
fundamental object $\mathscr{Z}$ in \eqref{Lie-setb} is replaced
by a FA element $Z$ as in \eqref{Lie-n-Lie}. It is sufficient to
note that, by the FI,
\begin{equation}
\begin{aligned}
ad_{\mathscr{X}\cdot\mathscr{Y}} Z &=
\sum_{i=1}^{n-1}[Y_1,\dots,Y_{i-1},[X_1,\dots,X_{n-1},Y_i],Y_{i+1},\dots,Y_{n-1},Z]
\\ & = [X_1,\dots,X_{n-1},[Y_1,\dots,Y_{n-1},Z]] -
[Y_1,\dots,Y_{n-1},[X_1,\dots,X_{n-1},Z]]\\ &=
ad_{\mathscr{X}}ad_{\mathscr{Y}}Z-ad_{\mathscr{Y}}ad_{\mathscr{X}}Z
\quad ,
\end{aligned}
\end{equation}
which completes the proof  $\square$

Note that eq.~\eqref{Lie-n-Lie} shows, by exchanging $\mathscr{X}$ and
$\mathscr{Y}$, that
$ad_{\mathscr{X}\cdot\mathscr{Y}}Z=-ad_{\mathscr{Y}\cdot\mathscr{X}}Z$
on any $Z\in \mathfrak{G}$, and hence that
\begin{equation}
\label{xy.skew} ad_{\mathscr{X}\cdot\mathscr{Y}} =
-ad_{\mathscr{Y}\cdot\mathscr{X}} \qquad \hbox{or, equivalently,}
\qquad (\mathscr{X}\cdot\mathscr{Y})\,\cdot =
-(\mathscr{Y}\cdot\mathscr{X})\,\cdot \quad,
\end{equation}
where the dots in the last expression should be noted, since
the composition of fundamental objects in eq.~\eqref{intro4}
is {\it not} commutative,  $\mathscr{X}\cdot\mathscr{Y}\not=
-\mathscr{Y}\cdot\mathscr{X}$. It follows from
eqs.~\eqref{xy.skew}, \eqref{Lie-n-Lie} that the
inner derivations $ad_{\mathscr{X}}\in \hbox{End}\,\mathfrak{G}$ of
a FA constitute an ordinary Lie algebra; therefore,
we have the following
\medskip

\noindent{\bf Proposition}
   Let $\mathfrak{G}$ a $n$-Lie algebra.  The inner drivations
$ad_{\mathscr{X}}$ associated to the fundamental objects $\mathscr{X}\in \wedge^{n-1}\mathfrak{G}$ determine
an ordinary Lie algebra, the {\it Lie algebra
associated with the FA}.
\medskip

\noindent
For instance, the Lie algebra associated to the $A_4$
euclidean  algebra (eq.~\eqref{simple} below for $n$=3 and no signs) is
$so(4)=so(3)\oplus so(3)$\footnote{Thus, the simple euclidean
$A_4$ determines $SO(4)$ as the gauge group of the
$A_4$-based BLG model. The fact that $SO(4)$ is not semisimple
was used \cite{Raam:08} to reformulate the BLG action with
no reference to $A_4$, with matter fields taking values in
the `bi-fundamental' representation of the gauge group
$SU(2)\otimes SU(2)$, and with the original gauge field replaced
by two $SU(2)$-gauge ones. The $\mathcal{N}$=6 model in
\cite{A-B-J-M:08}, which describes the low energy limit of the dynamics
of $N$ M2 branes, is constructed with scalar and fermion fields
taking values in the algebra of $U(N)\otimes U(N)$, with
a double set of gauge fields taking values in the adjoint,
and does not use a Filippov algebra structure. Its connection
with a (non fully skewsymmetric, see Sec. 6) three-bracket
structure was elucidated in  \cite{Ba-La:08}.}; for the
Lorentz case (one $\varepsilon=-1$),
an equally simple calculation leads to $so(1,3)$.

\section{On Kasymov's analogue of the Cartan-Killing form}

To prove that the cohomology groups that govern central extensions
and the infinitesimal deformations of semisimple $n$-Lie
algebras are trivial in analogy to the Lie algebra case, it would be
convinient to have that the form $k$ defined in (\ref{KasCK}),
viewed as a bilinear form on $\wedge^{n-1}\mathfrak{G}$
\begin{equation*}
 k:  \wedge^{n-1} \mathfrak{G} \times \wedge^{n-1} \mathfrak{G}
     \longrightarrow \mathbb{K} \; ,
\end{equation*}
be nondegenerate {\it i.e.}, that
$k(\mathscr{X},\mathscr{Y})=0\quad\forall\,\mathscr{Y}\in \wedge^{n-1}
\mathfrak{G} \Rightarrow \mathscr{X}=0$. In this way one could try
repeating for FAs the proof for Lie algebras. But we see immediately
that for $n\geq 3$ this is not so. Any semisimple $n$-Lie algebra is
the direct sum of its simple ideals \cite{Ling:93},
\begin{equation}
     \mathfrak{G} = \bigoplus_{\mathfrak{s}=1}^k
     \mathfrak{G}_{(\mathfrak{s})}
     = \mathfrak{G}_{(1)}\oplus \dots \oplus
     \mathfrak{G}_{(k)} \ .
\label{semisimple}
\end{equation}
As a consequence, the $n$-bracket $[\dots,X,\dots,Y,\dots]=0$
whenever $X$ and $Y$ belong to different simple ideals.
Then, if one considers, for instance, $\mathscr{X}=X_1\wedge\dots
\wedge X_{n-2}\wedge Y$, with $X_1,\dots ,X_{n-2}$ and $Y$ in
different ideals, it follows that $ad_{\mathscr{X}} \in
\textrm{End}\, \mathfrak{G}$ is identically zero, and so is
$k(\mathscr{X},\mathscr{Y})$ for any $\mathscr{Y}$ without
$\mathscr{X}$ itself being zero. In contrast, Kasymov criterion for
the semisimplicity of $n$-Lie algebras \cite{Kas:95a} establishes
that a FA is semisimple iff the $2(n-1)$-linear generalization $k$
is nondegenerate in the sense of eq.~\eqref{intro7}.

However, for {\it simple $n$-Lie algebras} the form $k$ is
nondegenerate on $\wedge^{n-1} \mathfrak{G}$. To show this,
we make use of the fact that a {\it real} simple $n$-Lie algebra
is one of the FA algebras given by \cite{Ling:93,Filippov}
\begin{equation}
\label{simple}
  [\textbf{e}_1\dots \hat{\textbf{e}}_i \dots \textbf{e}_{n+1}] =
  (-1)^{i+1} \varepsilon_i \textbf{e}_i \quad \textrm{or}
  \quad [\textbf{e}_{i_1}\dots \textbf{e}_{i_n}] = (-1)^n \sum^{n+1}_{i=1}
  \varepsilon_i {\epsilon_{i_1\dots i_n}}^i \textbf{e}_i \ ,
\end{equation}
where $\varepsilon_i=\pm 1$ (no sum over the $i$ of the $\varepsilon_i$
factors) just introduce signs\footnote{Note that
we might equally well have used the ${\epsilon_{i_1\dots i_n}}^i$
without signs $\varepsilon_i$ in the $r.h.s.$ by taking
${\epsilon_{i_1\dots i_n}}^i= \eta^{ij}\epsilon_{i_1\dots i_n j}$
where $\epsilon_{1\dots n (n+1)}=+1$ and $\eta$ is a
$(n+1)\times(n+1)$ diagonal metric with $+1$ and $-1$ in the places
indicated by the $\varepsilon_i$'s. We shall keep nevertheless the
customary $\varepsilon_i$ factors above as in {\it e.g.}
\cite{Filippov}.} that affect the different terms of the sum in $i$
and we have used Filippov's notation to denote the basis $\{\textbf{e}_i\}$
of $\mathfrak{G}$. Now, the bilinear form $k$ on $\wedge^{n-1}\mathfrak{G}$ is
determined by its values on a basis, so taking
$\mathscr{X}=(\textbf{e}_{i_1}, \dots , \textbf{e}_{i_{n-1}})$ and
$\mathscr{Y}=(\textbf{e}_{j_1}, \dots , \textbf{e}_{j_{n-1}})$, and
using eq.~\eqref{adZ},  the action of $ad_{\mathscr{X}}
ad_{\mathscr{Y}}$ on a basis vector $\textbf{e}_j$ is found to be
\begin{equation}
 ad_{\mathscr{X}} ad_{\mathscr{Y}} \,\textbf{e}_j = \sum_{l,s=1}^{n+1}
 \varepsilon_l \varepsilon_j \epsilon_{j_1 \dots j_{n-1} j}{}^l
 \epsilon_{i_1 \dots i_{n-1} l}{}^s \,\textbf{e}_s\ ,
\label{simple1}
\end{equation}
from which we deduce that the trace of $ad_{\mathscr{X}}
ad_{\mathscr{Y}}$ is given by
\begin{equation}
  k(\mathscr{X},\mathscr{Y}) = \sum_{l,s=1}^{n+1}
 \varepsilon_l \varepsilon_s \epsilon_{j_1 \dots j_{n-1} s}{}^l
 \epsilon_{i_1 \dots i_{n-1} l}{}^s\ .
\label{simple2}
\end{equation}
The matrix appearing on the $r.h.s.$ of (\ref{simple2}), seen as a
matrix $k_{(i_1\dotsi_{n-1})(j_1\dots j_{n-1})}$
with indices  $(i_1 \dots i_{n-1})$ and $(j_1 \dots j_{n-1})$
determined by the fundamental objects above, is clearly diagonal with non-zero
elements on the diagonal, for given $(i_1 \dots i_{n-1})$, the
factor $\epsilon_{i_1 \dots i_{n-1} l}{}^s$ fixes the remaining
indices $l$ and $s$ (and $\varepsilon_l \varepsilon_s $) so that
$(j_1, \dots, j_{n-1})$ has to be a reordering of the
 $(i_1 \dots i_{n-1})$ indices.
For this reason the form $k$ is diagonal with non-zero
elements in it and hence non-degenerate.

\section{Central extensions of $n$-Lie algebras}

\subsection{Cohomology and central extensions of Filippov
algebras}

Given a Filippov algebra $\mathfrak{G}$ with a $n$-bracket $[\dots
]$, we define a central extension $\widetilde{\mathfrak{G}}$ of
$\mathfrak{G}$ by adding a new, central, generator  $\Xi$ and
modifying the bracket as follows:
\begin{eqnarray}
  \label{cent-ext}
  & &  [\tilde{X}_{a_1},\dots ,\tilde{X}_{a_n}] :=
  f_{a_1\dots a_n}^b \,\tilde{X}_b + \alpha^1(X_1,\dots ,X_n) \Xi \ ,
\\
 \quad \quad & &  [\tilde{X}_1,\dots ,\tilde{X}_{n-1},\Xi]=0 \;,
 \nonumber
\end{eqnarray}
where the $f_{a_1\dots a_n}^b$ are the structure
constants of the unextended algebra $\mathfrak{G}$.
One may think of adding more than one central generator, but this
will not be needed here for the discussion. Clearly, $\alpha^1$ has
to be a $n$-linear and fully skewsymmetric map, $\alpha^1 \in
\wedge^{n-1} \mathfrak{G}^* \wedge \mathfrak{G}^*$, where
$\mathfrak{G}^*$  is the dual of $\mathfrak{G}$; it will be
identified with a {\it one}-cochain. Since the new bracket for the
$\tilde{X}_i$ has to satisfy the FI, this gives a condition on
$\alpha^1$. When one of the vectors involved is $\Xi$, the FI is
trivially satisfied; when $\Xi$ is absent it follows that
\begin{eqnarray}
\label{cohomology2}
 & &  [\tilde{X}_1,\dots ,\tilde{X}_{n-1},[\tilde{Y}_1,\dots ,\tilde{Y}_n]] = \nonumber\\
 & &  \quad \quad \quad \sum^n_{a=1} [\tilde{Y}_1,\dots , \tilde{Y}_{a-1},
 [\tilde{X}_1,\dots ,\tilde{X}_{n-1}, \tilde{Y}_a], \tilde{Y}_{a+1},\dots, \tilde{Y}_n]\; .
\end{eqnarray}
Using (\ref{cent-ext}) and the FI for the original Filippov algebra,
this implies
\begin{eqnarray}
 & &  \alpha^1(X_1,\dots ,X_{n-1},[Y_1,\dots ,Y_n]) - \nonumber\\
 & &  \quad \quad \quad \sum^n_{a=1} \alpha^1(Y_1,\dots , Y_{a-1},
 [X_1,\dots ,X_{n-1}, Y_a], Y_{a+1},\dots, Y_n)=0 \; .
\label{cohomology3}
\end{eqnarray}

\noindent This equation (with $Y_n=Z$), written as
$(\delta\alpha^1)(\mathscr{X},\mathscr{Y},Z)=0$, will provide below
the condition that characterizes $\alpha^1\in\wedge^{n-1}
\mathfrak{G}^* \wedge \mathfrak{G}^*$, $\alpha^1: \mathscr{X} \wedge
Z \mapsto \alpha^1(\mathscr{X}, Z)$ as a one-cocycle (for the
trivial action of $\mathfrak{G}$ on $\alpha$). It is seen now why
becomes natural to call $\alpha^1$ an one-cochain (rather than a
two-cochain, as it would be in the Lie algebra cohomology case) and
why we made the split
$\alpha^1\in\wedge^{n-1}\mathfrak{G}\wedge\mathfrak{G}$ explicit
rather than simply writing $\alpha^1\in\wedge^n\mathfrak{G}$: the
number of the fundamental objects in the arguments of a cochain
determines its order. As we shall see shortly,
an arbitrary $p$-cochain takes $p(n-1)+1$ arguments in $\mathfrak{G}$;
a zero-cochain is an element of $\mathfrak{G}^*$.

Let us now construct the cohomology complex relevant for central
extensions of FA. Since $\mathfrak{G}$ does not act on $\alpha^1
(\mathscr{X},Z)$, it will be the FA cohomology complex for
the trivial action. We define arbitrary $p$-cochains
as elements of $\wedge^{n-1}\mathfrak{G}^* \otimes \dots \otimes
\wedge^{n-1}\mathfrak{G}^* \wedge \mathfrak{G}^*$,
\begin{equation}
      \alpha^p: (\mathscr{X}_1, \dots ,\mathscr{X}_p, Z) \mapsto
      \alpha^p(\mathscr{X}_1, \dots ,\mathscr{X}_p, Z) \ ,
\label{cohomology4}
\end{equation}
where $\mathscr{X}_1,\dots ,\mathscr{X}_p$ are $p$ fundamental
objects. Condition (\ref{cohomology3}), which guarantees the
consistency of $\alpha^1$ in eq.~\eqref{cent-ext} with the FI
\eqref{cohomology2}, reads then
\begin{equation}
      (\delta \alpha^1)(\mathscr{X},\mathscr{Y},Z)=
      \alpha^1(\mathscr{X}, \mathscr{Y}\cdot Z)
      -\alpha^1(\mathscr{X}\cdot \mathscr{Y}, Z)  -
      \alpha^1(\mathscr{Y}, \mathscr{X}\cdot Z) = 0\ ,
\label{cohomology5}
\end{equation}
where $\mathscr{X}\cdot Z$ and $\mathscr{X}\cdot \mathscr{Y}$ were
defined in eqs. (\ref{adZ}) and (\ref{intro4}). It is now
straightforward to extend (\ref{cohomology5}) to a whole cohomology
complex; $\delta\alpha^p$ will be a $p+1$ cochain taking arguments
on one more fundamental object than $\alpha^p$. This is done by
means of the following

\medskip
\noindent {\bf Definition} ({\it FA cohomology complex
($C^\bullet(\mathfrak{G}), \delta)$
adapted to central extensions})

Let $\alpha^p\in \wedge^{n-1}\mathfrak{G}^* \otimes \dots \otimes
\wedge^{n-1}\mathfrak{G}^* \wedge \mathfrak{G}^*$ be a $p$-cochain on
a FA. The action of the coboundary operator $\delta$
on arbitrary $p$-cochains ($\alpha^p\in
C^p(\mathfrak{G})$) is given by (see \cite{AIP-B:97})
\begin{equation}
\label{n-cobop}
\begin{aligned}
(\delta\alpha)
(\mathscr{X}_1,\dots,\mathscr{X}_{p+1}, Z) & = \\
\sum_{1\leq i<j}^{p+1} (-1)^i \,& \alpha(\mathscr{X}_1,\dots,
\hat{\mathscr{X}}_i,\dots,\mathscr{X}_i\cdot\mathscr{X}_j,\dots,\mathscr{X}_{p+1},Z)\\
+  \sum_{i=1}^{p+1} (-1)^i\,  & \alpha (\mathscr{X}_1,\dots,
\hat{\mathscr{X}}_i,\dots,\mathscr{X}_{p+1}, \mathscr{X}_i \cdot Z) \; .
\end{aligned}
\end{equation}
The proof that $\delta^2=0$ is analogous to that for the Lie algebra
coboundary operator if we think of $\mathscr{X}\cdot\mathscr{Y}$
as a commutator, in which case eq.~\eqref{Lie-setb} plays the r\^ole
of a Jacobi identity (we shall come back to this point in Sec.~6).

The $p$-th cohomology groups are given by $H_0^p(\mathfrak{G})$
=$Z_0^p(\mathfrak{G})/B_0^p(\mathfrak{G})$, where
$Z_0^p(\mathfrak{G})$ is the group (for the natural addition of
cochains) of the $p$-cocycles, $Z_0^p(\mathfrak{G})= \{ \alpha^p\in
C^p(\mathfrak{G})| \delta \alpha =0\}$, and $B_0^p(\mathfrak{G})$
is the subgroup of the $p$-coboundaries, $B_0^p(\mathfrak{G})= \{
\alpha^p\in Z_0^p(\mathfrak{G})| \alpha^p=\delta \alpha^{p-1} ,\
\alpha^{p-1}\in C^{p-1}(\mathfrak{G}) \}$.
\medskip

A central extension is actually trivial if it is possible to find
new generators $\tilde{X}'\in\tilde{\mathfrak{G}}$ from the old
ones,
\begin{equation}
      \widetilde{X}'= \tilde{X}-\beta(X) \Xi \ ,
\label{cohomology7}
\end{equation}
where $\beta\in \mathfrak{G}^*$ is a zero-cochain, such that
they remove $\Xi$ from the $r.h.s.$ of eq.~\eqref{cent-ext},
\begin{eqnarray}
    [\widetilde{X}'_{a_1}, \dots , \widetilde{X}'_{a_n}] &=& f_{a_1\dots a_n}^b\widetilde{X}'_b \nonumber \\
    &=& \; f_{a_1\dots a_n}^b\tilde{X}_b - \beta([X_{a_1}, \dots , X_{a_n}])\Xi  \; .
\label{cohomology8}
\end{eqnarray}
Comparing the last term above with the original expression in
eq.~\eqref{cent-ext}, we conclude that a central trivial extension
is defined by a one-cochain of the form $\alpha^1$ such that
\begin{equation}
    \alpha^1(X_1, \dots , X_n ) = -\beta([X_1, \dots , X_n]) \; .
\label{cohomology10}
\end{equation}
This is tantamount to saying that the one-cocycle $\alpha^1$ is actually
the one-coboundary generated by the zero-cochain $\beta$,
$\alpha^1(\mathscr{X},Z)=(\delta\beta)(\mathscr{X},Z)$,
$(\delta\beta)(X_1, \dots ,X_{n-1},Z) = - \beta([X_1,\dots ,
X_{n-1},Z])$, as it is read from (\ref{n-cobop}). Clearly,
equivalent extensions correspond to one-cocycles that differ in a
coboundary. The different central extensions are thus characterized
by the elements of $H_0^1(\mathfrak{G})$, and the trivial extension
corresponds to the zero element of $H_0^1(\mathfrak{G})$.
This of course recovers the well
known Lie algebra cohomology result for $n=2$:
the second cohomology group $H^2_0(\mathfrak{g})$ for a Lie algebra
$\mathfrak{g}$ becomes the first one $H_0^1$ when $\mathfrak{g}$
is viewed as a FA $\mathfrak{G}$, since for $n=2$ the
fundamental objects are single elements $\mathscr{X}=X$ of
$\mathfrak{G}=\mathfrak{g}$.

\subsection{Triviality of the central extensions of
semisimple Filippov algebras}

We now show that all the central extensions of a semisimple $n$-Lie
algebra are trivial. To do so, we shall use the explicit form
of the simple $n$-Lie algebras in eq.~\eqref{simple}
\cite{Ling:93,Filippov} to prove  the
statement in this case first. Then, the decomposition
(\ref{semisimple}) will allow us to extend the result to
all semisimple $n$-Lie algebras.

As a previous example, and since the reasonings below may be
considered as a $n>2$ generalization of the $so(3)$ and
$so(1,2)$ Lie algebras, let us consider these $n=2$ cases first.
Their Lie algebra commutators may be jointly expressed as
$[X_i,X_j]=\varepsilon_k \epsilon_{ijk} X_k$, $\;i,j,k=1,2,3$. The
values $\varepsilon_1=\varepsilon_2=\varepsilon_3=1$ determine
$so(3)$,
\begin{equation}
    [X_1, X_2] = X_3 \; , \;  [X_2, X_3] = X_1   \; ,\; \quad [X_3, X_1] = X_2 \; ,
\end{equation}
whereas $so(1,2)$ corresponds to, say,
$\varepsilon_1=\varepsilon_2=1$, $\varepsilon_3=-1$,
\begin{equation}
    [X_1, X_2] = - X_3 \; , \; [X_2, X_3] = X_1 \;,\; [X_3, X_1] = X_2 \;.
\end{equation}
Since both $so(3)$ and $so(2,1)$ are simple, they do not have
non-trivial central extensions by Whitehead's Lemma. This is easy to
check directly. First, any two-cochain $\alpha^2$ is given by its
(skewsymmetric) coordinates $\alpha^2_{i_1i_2}= \alpha^2(X_{i_1},
X_{i_2})$. But this is also a two-cocycle since it satisfies
the two-cocycle condition $\alpha^2([X_{i_1},
X_{i_2}], X_{i_3} )- \alpha^2([X_{i_1}, X_{i_3}], X_{i_2}) +
\alpha^2([X_{i_2}, X_{i_3}], X_{i_1} )= 0$ (to check this, it suffices
to note that, since the antisymmetrization over
four indices is zero, $\epsilon_{[i_1 i_2 l}
\alpha^2_{i_3] l}\equiv 0$, which gives $\epsilon_{i_3 i_2 l}
\alpha^2_{i_3 l} - \epsilon_{i_1 i_3 l}\alpha^2_{i_2
l}+\epsilon_{i_2 i_3 l}\alpha^2_{i_1 l}=0$). But then $\alpha^2$ is a
two-coboundary, in fact the two-coboundary generated by the
one-cochain $\beta$, $\alpha^2=\delta\beta$, $(\delta\beta)(X_{i_1},
X_{i_2})= -\beta ([X_{i_1}, X_{i_2}])= \varepsilon_l
\epsilon_{i_1i_2 l} \beta_l$ with $\beta_l = \beta(X_l)$ so that
$\alpha^2_{ij}= -\varepsilon_l \epsilon_{i_1 i_2 l} \beta_l$. This
may always be satisfied with $\beta_l= \varepsilon_l \frac{1}{2}
{\epsilon_{ij}}_l \alpha^2_{ij}$. Note, however, that this type of
argument cannot be extended to other simple algebras, since
for all others $\hbox{dim}\,\mathfrak{g}$ will be larger than three,
and the structure constants will not be given in terms of the
three-dimensional skewsymmetric tensor.
\medskip

With this preliminary remark, ket us now move to the simple $n$-Lie
algebra case.

\medskip
\noindent
\textbf{Lemma}: Any one-cochain of a simple $n$-Lie algebra is a
one-coboundary (and thus a trivial one-cocycle).

{\it Proof}: Let $\alpha^1\in \wedge^{n-1} \mathfrak{G}^* \wedge
\mathfrak{G}^*$ be a one-cochain and $\mathfrak{G}$ simple. Given
a basis $\{\textbf{e}_i\}_{i=1}^{n+1}$ of $\mathfrak{G}$,
$\alpha^1$ is determined by its coordinates, $\alpha^1_{i_1\dots i_n}
=\alpha^1(\textbf{e}_{i_1},\dots ,\textbf{e}_{i_n})$. We now show
that, in fact, a one-cochain on a simple  $\mathfrak{G}$
is a one-coboundary {\it i.e.}, that there
exists a $\beta\in \mathfrak{G}^*$ such that
\begin{equation}
  \alpha^1_{i_1\dots i_n} = -\beta([\textbf{e}_{i_1}\dots
  \textbf{e}_{i_n}])=
  -\sum_{k=1}^{n+1} \varepsilon_k {\epsilon_{i_1\dots i_n}}^k  \beta_k
  \ ,
\label{central1}
\end{equation}
where $\beta_k=\beta(\textbf{e}_k)$. Indeed, given $\alpha^1$, the
zero-cochain $\beta$ given by
\begin{equation}
   \beta_k= -\frac{\varepsilon_k}{n!}\sum^{n+1}_{i_1\dots i_n=1}
   {\epsilon^{i_1\dots i_n}}_k \alpha^1_{i_1\dots i_n}
    \label{coboundaryextsimple}
\end{equation}
has the desired property (\ref{central1}):
\begin{eqnarray}
      -\beta([\textbf{e}_{i_1}\dots
  \textbf{e}_{i_n}])&=&
  -\sum_{k=1}^{n+1} {\epsilon_{i_1\dots i_n}}^k \varepsilon_k \beta_k
  \nonumber\\
  &=& \sum_{k=1}^{n+1} {\epsilon_{i_1\dots i_n}}^k
  \frac{\varepsilon_k^2}{n!} \sum^{n+1}_{j_1\dots j_n=1}
   {\epsilon^{j_1\dots j_n}}_k \alpha^1_{j_1\dots j_n} \nonumber\\
   &=& \frac{1}{n!} \sum^{n+1}_{j_1\dots j_n=1}
   \epsilon^{j_1\dots j_n}_{i_1\dots i_n} \alpha^1_{j_1\dots j_n}
   = \alpha^1_{i_1\dots i_n} \; ,
\label{central2}
\end{eqnarray}
which proves the lemma $\square$
\medskip

Let now $\mathfrak{G}$ be a semisimple FA, and
(\ref{semisimple}) the splitting in its simple components. First, we
establish the following simple
\medskip

\noindent {\bf Lemma}: Let $\alpha^1\in \wedge^{n-1} \mathfrak{G}^*
\wedge \mathfrak{G}^*$ be a one-cocycle on a semisimple $n$-Lie
algebra for the $n$-Lie algebra trivial action cohomology defined by
eq.~\eqref{n-cobop}. Then, $\alpha(X_1,\dots,X_{n-2},Y,Z)=0$ if $Y$
and $Z$ belong to different ideals.

{\it Proof}:
Let the different simple ideals be labelled
by small gothic letters $\mathfrak{s}$, $\mathfrak{t}$, etc. Let $Z\in
\mathfrak{G}_{(\mathfrak{s})}$ and $Y\in
\mathfrak{G}_{(\mathfrak{t})}$, $\mathfrak{s}\neq \mathfrak{t}$.
Since $\mathfrak{G}_{(\mathfrak{s})}$ is simple, there exist
$Z_1,\dots,Z_n \in \mathfrak{G}_{(\mathfrak{s})}$ such that
$[Z_1,\dots,Z_n]=Z$. We can now use this fact and the cocycle
condition in eq.~\eqref{cohomology3} to obtain
\begin{eqnarray}
 & & \alpha(X_1,\dots,X_{n-2},Y,Z) = \alpha^1(X_1,\dots,X_{n-2},Y,[Z_1,\dots,Z_n])
  \nonumber\\
 & & \quad\quad \quad \quad  = \sum_{k=1}^n \alpha^1(Z_1,\dots
   [X_1,\dots,X_{n-2},Y,Z_k],\dots,Z_n)=0
\label{central3}
\end{eqnarray}
because $[X_1,\dots ,X_{n-2},Y,Z]=0$ since $Y\in
\mathfrak{G}_{(\mathfrak{t})}$ and $Z\in
\mathfrak{G}{(\mathfrak{s})}$ and $\mathfrak{s}\neq \mathfrak{t}\;$
 $\square$
\medskip

Using this lemma, we can now prove the main result of this
section:

\medskip
\noindent
{\bf Theorem} All central extensions of semisimple $n$-Lie algebras
are trivial.

{\it Proof}: Let $\alpha^1\in \wedge^{n-1} \mathfrak{G}^* \wedge
\mathfrak{G}^*$ be a one-cocycle in the FA cohomology for the trivial
action, and let $\mathfrak{G}$  be semisimple.
Then the theorem follows if $\alpha^1$ is a one-coboundary.

All $X\in \mathfrak{G}$ can be split in a unique way in the form
$X=\sum_{\mathfrak{s}=1}^k X_{(\mathfrak{s})}$, where
$X_{(\mathfrak{s})}$ is the component of $X$ in the simple ideal
labelled by $\mathfrak{s}$. Then we have
\begin{equation}
  \alpha^1(X_1,\dots,X_n) = \sum_{\mathfrak{s}_1\dots
  \mathfrak{s}_n=1}^k
  \alpha^1(X_{1(\mathfrak{s}_1)},\dots,X_{n(\mathfrak{s}_n)}) =
  \sum_{\mathfrak{s}=1}^k
  \alpha^1(X_{1(\mathfrak{s})},\dots,X_{n(\mathfrak{s})}) \; ,
\label{central4}
\end{equation}
where the last equality is due to the previous lemma. Every term in
the above expression defines a cochain on the simple $n$-Lie algebra
$\mathfrak{G}_{(\mathfrak{s})}$, so they are coboundaries {\it i.e.}
there exist $\beta_{(\mathfrak{s})} \in
\mathfrak{G}_{(\mathfrak{s})}^*$ such that
$\alpha^1(X_{1(\mathfrak{s})},\dots,X_{n(\mathfrak{s})}) =
-\beta_{(\mathfrak{s})}([X_{1(\mathfrak{s})},\dots,X_{n(\mathfrak{s})}])$.
This means that there is a zero-cochain $\beta\in \mathfrak{G}^*$,
\begin{equation}
  \beta(X) = \sum_{\mathfrak{s}=1}^k \beta_{(\mathfrak{s})}(X_{(\mathfrak{s})}) \ ,
\label{central5}
\end{equation}
that generates $\alpha^1$:
\begin{eqnarray}
  -\beta([X_1,\dots , X_n]) &=& -\beta\left(\sum^k_{(\mathfrak{s}_1)\dots (\mathfrak{s}_n)=1}
 [X_{1(\mathfrak{s}_1)},\dots,X_{n(\mathfrak{s}_n)}]\right) \nonumber\\
 &=& -\beta \left( \sum^k_{(\mathfrak{s})=1} [X_{1(\mathfrak{s})},\dots,X_{n(\mathfrak{s})}]\right) =
 -\sum_{\mathfrak{s}=1}^k \beta_{(\mathfrak{s})}([X_{1(\mathfrak{s})},\dots,X_{n(\mathfrak{s})}]) \nonumber\\
 &=& \sum_{\mathfrak{s}=1}^k \alpha^1(X_{1(\mathfrak{s})},\dots,X_{n(\mathfrak{s})})= \alpha^1(X_1,\dots ,
 X_n)\ ,
\label{central6}
\end{eqnarray}
which concludes the proof $\square$

\section{Infinitesimal deformations of $n$-Lie algebras}

\subsection{Cohomology complex adapted to deformations of
Filippov algebras}

In contrast with algebra extensions, which add one or more generators,
deformations \cite{Gers:63,Nij-Rich:67} do not increase the
dimension of the algebra. The infinitesimally deformed $n$-bracket
$[\dots ]_t$ may be written in terms of the original one $[\dots ]$
as follows:
\begin{equation}
\label{ntacohomology1}
    [X_1 ,\dots , X_n]_t = [X_1 ,\dots , X_n] +t \alpha^1(X_1 ,\dots ,
    X_n) \ ,
\end{equation}
where now $\alpha^1$ is $\mathfrak{G}$-valued (the added term
must belong to $\mathfrak{G}$), and $t$ is the
parameter of the infinitesimal deformation
\cite{Gers:63,Nij-Rich:67}. The one-cocycle condition for the
deformation problem appears when the deformed $n$-bracket in
(\ref{ntacohomology1}) is made to satisfy the FI so that the
deformed bracket does define a FA at order $t$
\begin{eqnarray}
    & & [X_1 ,\dots , X_{n-1}, [Y_1,\dots , Y_n]_t]_t\nonumber \\
    & & \quad\quad \quad= \sum_{a=1}^n [Y_1 ,\dots , Y_{a-1},
    [X_1 ,\dots , X_{n-1}, Y_a]_t, Y_{a+1}, \dots , Y_n]_t  \; .
\label{ntacohomology2}
\end{eqnarray}
In terms of fundamental objects (and setting $Y_n=Z$) the above
condition reads
\begin{equation}
\label{def-3JI}
 [\mathscr{X}, (\mathscr{Y}\cdot Z)_t]_t= [(\mathscr{X}\cdot\mathscr{Y})_t, Z]_t
 +[\mathscr{Y},(\mathscr{X}\cdot Z)_t]_t
\end{equation}
({\it cf} eq.~\eqref{Lie-n-Lie}). Eq.~\eqref{ntacohomology2}
implies, keeping only terms linear in $t$,
\begin{eqnarray}
    & & [X_1 ,\dots , X_{n-1}, \alpha^1(Y_1,\dots , Y_n)]
    +\alpha^1(X_1 ,\dots , X_{n-1}, [Y_1,\dots , Y_n])\nonumber\\
    & & \quad= \sum_{a=1}^n [Y_1 ,\dots , Y_{a-1},
    \alpha^1(X_1 ,\dots , X_{n-1}, Y_a), Y_{a+1}, \dots ,
    Y_n]\nonumber\\
    & & \quad +
    \sum_{a=1}^n \alpha^1(Y_1 ,\dots , Y_{a-1},
    [X_1 ,\dots , X_{n-1}, Y_a], Y_{a+1}, \dots , Y_n)\; .
\label{ntacohomology3}
\end{eqnarray}

This expression may be read as the one-cocycle
condition $\delta\alpha^1=0$ for the $\mathfrak{G}$-valued cochain
$\alpha^1$. In terms of the fundamental objects it may be written,
setting again $Y_n=Z$, as
\begin{equation}
\label{def-1-coch-b}
\begin{aligned}
(\delta\alpha^1)(\mathscr{X}, \mathscr{Y}, Z) = ad_{\mathscr{X}}&
\alpha^1(\mathscr{Y}, Z) -ad_{\mathscr{Y}}\alpha^1(\mathscr{X}, Z)
  -(\alpha^1(\mathscr{X}, \quad )\cdot \mathscr{Y})\cdot Z \\
 -\alpha^1 (\mathscr{X} & \cdot\mathscr{Y}, Z)  -\alpha^1(\mathscr{Y}, \mathscr{X}\cdot Z)
 + \alpha^1(\mathscr{X}, \mathscr{Y}\cdot Z) = 0 \quad ,
 \end{aligned}
\end{equation}
where, for instance for $n=3$, the term $\alpha^1(\mathscr{X},\quad)\cdot
\mathscr{Y}$ above is the fundamental object defined by
\begin{equation}
\label{dosf}
\begin{aligned}
 \alpha^1(\mathscr{X},\quad)\cdot \mathscr{Y} :=&
 (\alpha^1(\mathscr{X},\quad)\cdot Y_1,\,Y_2) \,+\, (Y_1,\,\alpha^1(\mathscr{X},\quad)\cdot Y_2)\\
 =& (\alpha^1(\mathscr{X},Y_1), Y_2) \,+\, (Y_1,\alpha^1(\mathscr{X},Y_2)) \quad ,\quad
[\,\alpha^1(\mathscr{X},\quad)\cdot Y_i
:=\alpha^1(\mathscr{X},Y_i)\,] \quad .
\end{aligned}
\end{equation}
The general action of the coboundary operator on an arbitrary
$p$-cochain will be given in eq. (\ref{adcoho}) below; we notice
at this stage that expression (\ref{def-1-coch-b}) involves {\it
both} the left and right actions of $\mathfrak{G}$ on $\alpha^1$.

An infinitesimal deformation is trivial if there exists a
redefinition $X'_i=X_i-t\beta(X_i)$, for some $\mathfrak{G}$-valued
zero-cochain $\beta$ that removes the deformating term in
eq.~\eqref{ntacohomology1}. If so, the first order deformed bracket
in terms of the new, primed generators reads
\begin{eqnarray}
\label{ntacohomology4}
    [{X'}_1, \dots , {X'}_n]_t & = &  [X_1, \dots , X_n]'
      \nonumber  \\
     & \equiv &
    [X_1, \dots , X_n]- t\beta([X_1, \dots , X_n]) \; .
\end{eqnarray}
But, again keeping terms up to order $t$ only, we find that
the $l.h.s.$ above gives
\begin{eqnarray}
   [{X'}_1 ,\dots , {X'}_n]_t &=& [X_1,\dots ,X_n]_t \nonumber \\
    &&-t \sum_{a=1}^n [X_1 ,\dots , X_{a-1},
    \beta(X_a), X_{a+1}, \dots , X_n]_t\nonumber\\
    & & \nonumber\\
   &=&[X_1,\dots ,X_n] + t \alpha^1(X_1,\dots ,X_n)
      \nonumber  \\
     &&-t \sum_{a=1}^n [X_1 ,\dots , X_{a-1},
    \beta(X_a), X_{a+1}, \dots , X_n]\; .
\label{ntacohomology5}
\end{eqnarray}
Therefore, the infinitesimal deformation given by the
$\mathfrak{G}$-valued one-cocycle $\alpha^1$ is trivial if there
exists a $\mathfrak{G}$-valued zero-cochain $\beta$ such that
$\alpha^1=\delta\beta$ where
\begin{equation}
    (\delta\beta)(X_1, \dots , X_n) : =
    -\beta([X_1, \dots , X_n]) +\sum_{a=1}^n [X_1 ,\dots , X_{a-1},
    \beta(X_a), X_{a+1}, \dots ,
    X_n] \ .
\label{ntacohomology6}
\end{equation}
This is the expression that defines the one-coboundary generated by
$\beta$ in the FA cohomology induced by the deformation problem (eq.
(\ref{adcoho}) below for $\alpha^0=\beta$).

The expression of the one-coboundary may again be formulated in
terms of the fundamental objects $\mathscr{X}$, as in the central
extensions case, which will allow us to generalize the action of the
coboundary operator $\delta$ on an arbitary cochain $\alpha^p\in
C^p(\mathfrak{G},\mathfrak{G})$. Explicitly we find, relabelling
the zero-cochain $\beta$ as $\alpha^0$,
\begin{equation}
\label{gval-one-cob}
 (\delta\alpha^0)(\mathscr{X},Z)= \mathscr{X}\cdot \alpha^0(Z)
 -\alpha^0(\mathscr{X}\cdot Z)
 + (\alpha^0(\quad)\cdot \mathscr{X})\cdot Z \quad .
\end{equation}
As for Lie algebras, it is the
characteristic identity, here the FI, that is responsible for
the structure of the whole cohomology complex. This leads to
a generalization of the Lie algebra cohomology relative
to the adjoint action, and it is given by the following
\medskip

\noindent {\bf Definition} ({\it Cohomology complex
$(C^\bullet_{ad}(\mathfrak{G},\mathfrak{G}),\,\delta)$ for
deformations of FA})

Let $\alpha^p \in C^p(\mathfrak{G},\mathfrak{G})$ a $\mathfrak{G}$-valued
$p$-cochain, $\alpha^p: \wedge^{(n-1)}\mathfrak{G}\otimes
\mathop{\cdots}\limits^p\otimes\wedge^{(n-1)}\mathfrak{G}\wedge\mathfrak{G} \rightarrow \mathfrak{G}$.
The action of the coboundary operator is given by
\begin{equation}
\begin{aligned}
\label{adcoho}
(\delta\alpha^p) &
(\mathscr{X}_1,\dots,\mathscr{X}_p,\mathscr{X}_{p+1},Z)= \\
&\sum_{1\leq j<k}^{p+1} (-1)^j
\alpha^p(\mathscr{X}_1,\dots,\widehat{\mathscr{X}_j},\dots,\mathscr{X}_{k-1},
\mathscr{X}_j\cdot
\mathscr{X}_k,\mathscr{X}_{k+1},\dots,\mathscr{X}_{p+1}, Z)\\
+& \sum_{j=1}^{p+1} (-1)^j \alpha^p
(\mathscr{X}_1,\dots,\widehat{\mathscr{X}_j},\dots,\mathscr{X}_{p+1},\mathscr{X}_j\cdot
 Z) \\
 + &\sum_{j=1}^{p+1} (-1)^{j+1} \mathscr{X}_j \cdot
 \alpha^p(\mathscr{X}_1,\dots,\widehat{\mathscr{X}_j}\dots,\mathscr{X}_{p+1},
 Z) \\
 +&(-1)^{p}
 (\alpha^p(\mathscr{X}_1,\dots,\mathscr{X}_p\,,\quad)\cdot
 \mathscr{X}_{p+1})\cdot Z
\end{aligned}
\end{equation}
where, in the last term,
\begin{equation}
\label{p-coc-ftal}
 \alpha^p(\mathscr{X}_1,\dots,\mathscr{X}_p\,,\quad)\cdot
 \mathscr{Y}=
 \sum_{i=1}^{n-1}
 (Y_1,\dots,\alpha^p(\mathscr{X}_1,\dots,\mathscr{X}_p,Y_i),\dots,Y_{n-1})
 \quad .
\end{equation}
\medskip

The above equations define
($C^\bullet_{ad}(\mathfrak{G},\mathfrak{G}),\delta$) as induced by
the deformation problem. It is seen above that both left and right
`$ad$' actions enter into the definition of the coboundary operator.
The above cohomological
complex may be seen  essentially equivalent to the
one adapted to deformations introduced by Gautheron \cite{Gau:96}
for the Nambu-Poisson algebras; see also
\cite{Da-Tak:97, Tak:94, Rot:05}.
Since $H_{ad}^p(\mathfrak{G},\mathfrak{G}) =
Z_{ad}^p(\mathfrak{G},\mathfrak{G})/B_{ad}^p(\mathfrak{G},\mathfrak{G})$
it follows that the infinitesimal deformations of Filippov algebras
are governed by $H_{ad}^1(\mathfrak{G},\mathfrak{G})$.

   It is not difficult to check that if one moves to the next order
in the deformation of $\mathfrak{G}$ by adding
$t^2 \alpha^{(2)}(\mathscr{X},X_n)$
to the $r.h.s.$ of eq.~\eqref{ntacohomology1}, where the superindex
two in $\alpha^{(2)}$ refers to the order of the deformation, the FI
imposes on the one-cochain $\alpha^{(2)}$ a condition of the form
$\gamma(\mathscr{X},\mathscr{Y},Z)=
(\delta \alpha^{(2)})(\mathscr{X},\mathscr{Y},Z)$, where
$\gamma(\mathscr{X},\mathscr{Y},Z)$ is a two-cocycle which
is expressed in terms of $\alpha^1$. Therefore, if
$H^2_{ad}(\mathfrak{G},\mathfrak{G})\not=0$, there is an
obstruction that prevents extending the infinitesimal
deformation to higher orders. As a result,
we have the following
\medskip

\noindent {\bf Theorem} ({\it Deformations of FA algebras})

Let $\mathfrak{G}$ be a FA. The first cohomology group
$H^1_{ad}(\mathfrak{G},\mathfrak{G})$ for the above complex governs
the infinitesimal deformations of $\mathfrak{G}$. The triviality of
this cohomology group, $H_{ad}^1=0$, is a sufficient condition for
the rigidity of a FA. The obstruction to expanding an
infinitesimal deformation is given by the non-vanishing of
the second cohomology group, $H^2_{ad}(\mathfrak{G},\mathfrak{G})\not=0$.

{\it Proof}: Contained above $\square$
\medskip

For $n=2$, and reverting to the notation that characterizes the order
of cochains and cocycles by the number of their Lie algebra
arguments, the above theorem recovers the standard result~\cite{Gers:63, Nij-Rich:67}
for Lie algebras: the infinitesimal
deformations are characterized by $H^2_{ad}(\mathfrak{g},\mathfrak{g})$
and the obstructions to go to higher orders are in the non-vanishing
elements of third cohomology group $H^3_{ad}(\mathfrak{g},\mathfrak{g})$.

\subsection{Rigidity of semisimple Filippov algebras}

In this section we  prove that semisimple $n$-Lie algebras are also
rigid. The pattern of the proof is similar to the case of the
central extensions: first we present the proof for simple FAs and
then extend it to the semisimple $n$-Lie algebras. In contrast with
the central extensions problem, it is not true now
that every one-cochain of a simple $n$-Lie algebra is a
coboundary, so it is necessary to characterize the cocycles for
simple $n$-Lie algebras, eq.~(\ref{simple}). Taking a basis
of $\mathfrak{G}$ $\{\textbf{e}_i\}_{i=1}^{n+1}$, a one-cochain is
defined by
\begin{equation}
   \alpha^1(\textbf{e}_{i_1},\dots,\textbf{e}_{i_n})=\sum_{j=1}^{n+1}
   {(\alpha^1)^j}_{i_1\dots i_n}\textbf{e}_j\ ,
\label{deformation2}
\end{equation}
in terms of its coordinates ${(\alpha^1)^j}_{i_1\dots i_n}$. It will
turn out convenient to define new, dual quantities that are easier
to manipulate:
\begin{equation}
  (\alpha^1)^{ji} := \frac{1}{n!} \sum_{i_1\dots i_n=1}^{n+1} \epsilon^{i_1\dots i_n i}
 {(\alpha^1)^j}_{i_1\dots i_n}\ ,\quad {(\alpha^1)^j}_{i_1\dots
 i_n}=  \sum_{i=1}^{n+1} \epsilon_{i_1\dots i_n i}
 (\alpha^1)^{ji}\ .
\label{deformation3}
\end{equation}
Using them we now prove the following

\medskip

{\bf Proposition}\label{prop} Let $\alpha^1$ be a
$\mathfrak{G}$-valued one-cochain of a simple $n$-Lie algebra with
coordinates ${(\alpha^1)^j}_{i_1\dots i_n}$. Then, $\alpha^1$ is a
one-cocycle for the above cohomology complex {\it iff}
$(\alpha^1)^{ji}$ is symmetric, $(\alpha^1)^{ji}=(\alpha^1)^{ij}$.

\medskip

{\it Proof}:
 First, we write the one-cocycle condition
(\ref{ntacohomology3}) in terms of basis elements of
$\mathfrak{G}$,
\begin{eqnarray}
0&=& [\textbf{e}_{i_1},\dots,\textbf{e}_{i_{n-1}},
\alpha^1(\textbf{e}_{j_1},\dots,\textbf{e}_{j_{n-1}},\textbf{e}_k)]
\nonumber\\
& & +
\alpha^1(\textbf{e}_{i_1},\dots,\textbf{e}_{i_{n-1}},[\textbf{e}_{j_1},
\dots,\textbf{e}_{j_{n-1}},\textbf{e}_k])\nonumber\\
 & & - \sum_{a=1}^{n-1}
\alpha^1(\textbf{e}_{j_1},\dots, \textbf{e}_{j_{a-1}},
[\textbf{e}_{i_1},\dots, \textbf{e}_{i_{n-1}},\textbf{e}_{j_a}],
\textbf{e}_{j_{a+1}},\dots ,\textbf{e}_{j_{n-1}},\textbf{e}_k)
\nonumber\\
& & -\alpha^1(\textbf{e}_{j_1},\dots,\textbf{e}_{j_{n-1}},
[\textbf{e}_{i_1},\dots,\textbf{e}_{i_{n-1}},\textbf{e}_k])
\nonumber\\
& & - \sum_{a=1}^{n-1} [\textbf{e}_{j_1},\dots,
\textbf{e}_{j_{a-1}},\alpha^1(\textbf{e}_{i_1},\dots,\textbf{e}_{i_{n-1}},
\textbf{e}_{j_a}), \textbf{e}_{j_{a+1}},\dots ,
\textbf{e}_{j_{n-1}},\textbf{e}_k]\nonumber\\ & & -
[\textbf{e}_{j_1},\dots,\textbf{e}_{j_{n-1}},
\alpha^1(\textbf{e}_{i_1},\dots,\textbf{e}_{i_{n-1}},
\textbf{e}_k)] \ . \label{deformation4}
\end{eqnarray}
Using the commutation relations (\ref{simple}) of the simple
algebra and that
\begin{equation}
 \alpha^1(\textbf{e}_{i_1},\dots,\textbf{e}_{i_{n-1}},\textbf{e}_k)
 = \sum_{l=1}^{n+1} \textbf{e}_l {(\alpha^1)^l}_{i_1\dots
 i_{n-1}k} = \sum_{r,s=1}^{n+1} \textbf{e}_r \,\epsilon_{i_1 \dots i_{n-1}ks}
 (\alpha^1)^{rs}
\label{deformation4a}
\end{equation}
by eq. (\ref{deformation3}), eq. (\ref{deformation4}) gives
\begin{eqnarray}
0&=& \sum_{s,l=1}^{n+1} \Bigg( \varepsilon_r {\epsilon_{i_1\dots
i_{n-1}s}}^r\epsilon_{j_1\dots j_{n-1}kl} (\alpha^1)^{sl}
\nonumber\\
& & +  \varepsilon_ s \epsilon_{i_1\dots i_{n-1} sl}
{\epsilon_{j_1\dots j_{n-1}k}}^s (\alpha^1)^{rl}\nonumber\\
& &- \sum_{a=1}^{n-1} \varepsilon_s {\epsilon_{i_1\dots i_{n-1}j_a
}}^s \epsilon_{j_1\dots j_{a-1}sj_{a+1}\dots j_{n-1}kl}
(\alpha^1)^{rl}  \nonumber\\
& & - \varepsilon_ s \epsilon_{j_1\dots j_{n-1}sl}
{\epsilon_{i_1\dots i_{n-1}k}}^s (\alpha^1)^{rl}  \nonumber\\
& & -  \sum_{a=1}^{n-1} \varepsilon_r   {\epsilon_{j_1\dots
j_{a-1}sj_{a+1}\dots j_{n-1}k}}^r \epsilon_{i_1\dots i_{n-1}j_a l}
(\alpha^1)^{sl} \nonumber\\
& &- \varepsilon_r {\epsilon_{j_1\dots j_{n-1}s}}^r
\epsilon_{i_1\dots i_{n-1}k l} (\alpha^1)^{sl} \Bigg)  \
.\label{deformation5}
\end{eqnarray}
Since this expression translates the condition that
$\delta\alpha^1(\mathscr{X},\mathscr{Y},\textbf{e}_k) = 0$, where
$\mathscr{X} \in \wedge^{n-1} \mathfrak{G}$, $\mathscr{Y} \in
\wedge^{n-1} \mathfrak{G}$ with
$\mathscr{X}=(\textbf{e}_{i_1},\dots,\textbf{e}_{i_{n-1}})$,
$\mathscr{Y} = (\textbf{e}_{j_1},\dots,\textbf{e}_{j_{n-1}})$ , it
follows that it must be antisymmetric in the indices $i$ and also in
the indices $j$ and, indeed, it may be checked explicitly in
eq.~\eqref{deformation5} that this is so. Using the antisymmetry in
$i_1,\dots , i_{n-1}$ and in $j_1,\dots ,j_{n-1}$, an equivalent
expression is obtained by contracting with $\epsilon^{i_1\dots
i_{n+1}}$ and $\epsilon^{j_1\dots j_{n+1}}$,
\begin{equation}
   \sum^{n+1}_{s,l=1}(\epsilon^{i_ni_{n+1}}_{sr}\epsilon^{j_nj_{n+1}}_{kl}
   - \epsilon^{i_ni_{n+1}}_{lr}\epsilon^{j_nj_{n+1}}_{ks}-
   \epsilon^{i_ni_{n+1}}_{sl}\epsilon^{j_nj_{n+1}}_{kr})
   (\alpha^1)^{sl} = 0 \; .
 \label{deformation6}
\end{equation}
Clearly if $(\alpha^1)^{sl}$ is symmetric then (\ref{deformation6})
is satisfied. Conversely, contracting (\ref{deformation6}) with
$\delta_{j_{n+1}}^k \delta_{i_{n+1}}^r$ we obtain
\begin{equation}
   (n-n^2) \left( (\alpha^1)^{i_nj_n} - (\alpha^1)^{j_ni_n} \right) = 0\  ,
 \label{deformation7}
\end{equation}
which means that (\ref{deformation6}) holds if and only if
$(\alpha^1)^{ij}$ is symmetric $\square$

Using the previous Proposition it now follows that the  simple
Filippov $n$-Lie algebras are stable in the sense of Gerstenhaber
\cite{Gers:63,Nij-Rich:67}

\medskip
\noindent
{\bf Theorem}
All infinitesimal deformations of a simple $n$-Lie algebra are
trivial and therefore simple FAs are rigid.
\medskip

{\it Proof:}  We have to show that if $\alpha^1$ is a one-cocycle it
is a trivial one. Let the one-cochain be characterized by
$(\alpha^1)^{ij}$ as in eq. (\ref{deformation3}). In order to
express the one-coboundary condition in terms of $(\alpha^1)^{ij}$,
we rewrite the coboundary condition
(\ref{ntacohomology6}) in the basis $\{ \textbf{e}_i\}$,
\begin{eqnarray}
  & & \alpha^1(\textbf{e}_{i_1},\dots ,\textbf{e}_{i_n}) = \delta
   \beta(\textbf{e}_{i_1},\dots ,\textbf{e}_{i_n}) =-
   \beta([\textbf{e}_{i_1},\dots ,\textbf{e}_{i_n}]) \nonumber\\
   & & \quad \quad\quad\quad + \sum_{a=1}^n [ \textbf{e}_{i_1},\dots ,\textbf{e}_{i_{a-1}},
    \beta(\textbf{e}_{i_a}),
   \textbf{e}_{i_{a+1}}, \dots , \textbf{e}_{i_n}]\ ,
 \label{deformation8}
\end{eqnarray}
which implies, after using
$\beta(\textbf{e}_j):=\textbf{e}_i{\beta^i}_j $, $i=1,\dots , n+1$,
that
\begin{eqnarray}
   {(\alpha^1)^r}_{i_1\dots i_n} &=&  - (-1)^n \sum^{n+1}_{s=1}
   \varepsilon_s {\epsilon_{i_1\dots i_n}}^s {\beta^r}_s \nonumber
   \\
   & & + (-1)^n \sum^n_{a=1} \sum^{n+1}_{s=1} \varepsilon_r
   {\epsilon_{i_1\dots i_{a-1} s i_{a+1} \dots i_n}}^r
   {\beta^s}_{i_a}\ .
 \label{deformation9}
\end{eqnarray}
If we now contract this equation with $\epsilon^{i_1\dots i_nk}$ we
find that the coboundary condition may be rewritten as
$(\alpha^1)^{rk}=(\delta \beta)^{rk}$ with
\begin{equation}
   (\delta\beta)^{rk} = -(-1)^n(\varepsilon_k \beta^{rk} + \varepsilon_r \beta^{kr})
   + (-1)^n \sum^{n+1}_{s=1} {\beta^s}_s \varepsilon_r \delta^{rk}
   \ .
 \label{deformation10}
 \end{equation}
Let $\alpha^1$ now be a cocycle. Then, it is generated by the zero-cochain
$\beta$ given by
\begin{equation}
   (\beta)^{jk} = -\frac{(-1)^n}{2} \left[ \varepsilon_k
   (\alpha^1)^{jk}- \frac{1}{n-1} \sum^{n+1}_{s=1} \varepsilon_s
   {(\alpha^1)^s}_s \delta^{jk} \right]  \quad.
 \label{deformation101}
 \end{equation}
Indeed, inserting (\ref{deformation101}) into
the $r.h.s$ of (\ref{deformation10}) we get
\begin{eqnarray}
   (\delta \beta)^{jk} & = & \frac{1}{2}\left( (\alpha^1)^{jk} - \frac{1}{n-1}
   \varepsilon_k \delta^{jk} \sum^{n+1}_{s=1} \varepsilon_s
   {(\alpha^1)^s}_s \right.\nonumber\\
   & &+ \left.(\alpha^1)^{kj} - \frac{1}{n-1}
   \varepsilon_j \delta^{jk} \sum^{n+1}_{s=1} \varepsilon_s
   {(\alpha^1)^s}_s \right) \nonumber\\
   & & - \frac{1}{2} \varepsilon_j \delta^{jk} \left( \sum^{n+1}_{s=1} \varepsilon_s
   {(\alpha^1)^s}_s - \frac{n+1}{n-1} \sum^{n+1}_{s=1} \varepsilon_s
   {(\alpha^1)^s}_s\right)\nonumber \\
   &=&  (\alpha^1)^{jk}\ ,
 \label{deformation11}
\end{eqnarray}
since all sums in the expression cancel among themselves and, because
$\alpha^1$ is a one-cocycle, $\alpha^{jk}$ is symmetric by the
previous Proposition. Therefore every one-cocycle of a simple
$n$-Lie algebra is trivial and simple FAs are rigid $\square$
\medskip

Again, this result can be extended to semisimple $n$-Lie algebras.
By eq.~\eqref{semisimple} every $X\in \mathfrak{G}$ can
be written as a sum $X=\sum_{\mathfrak{s}=1}^l X_{(\mathfrak{s})}$
of components in the different simple ideals, $X_{(\mathfrak{s})}
\in \mathfrak{G}_{(\mathfrak{s})}$. Then, the result of the action
of $\alpha^1$ on $n$ vectors $X_1,\dots ,X_n \in \mathfrak{G}$ may
be written as follows:
\begin{eqnarray}
    \alpha^1(X_1,\dots ,X_n) &=& \alpha^1 \left(
    \sum^k_{\mathfrak{s}_1 =1} X_{1(\mathfrak{s}_1)}, \dots ,
    \sum^k_{\mathfrak{s}_n =1} X_{n(\mathfrak{s}_n)} \right)
    \nonumber\\
    &=& \sum^k_{\mathfrak{s}_1 \dots \mathfrak{s}_n =1}
    \alpha^1(X_{1(\mathfrak{s}_1)}, \dots ,
    X_{n(\mathfrak{s}_n)})\nonumber \\
    &=& \sum^k_{\mathfrak{s}_1 \dots \mathfrak{s}_n =1}
    \sum^k_{\mathfrak{t}=1}
    (\alpha^1)^{(\mathfrak{t})}(X_{1(\mathfrak{s}_1)}, \dots ,
    X_{n(\mathfrak{s}_n)}) \ .
 \label{deformation12}
\end{eqnarray}
As a result, a one-cochain for a semisimple FA $\mathfrak{G}$ is
determined once the components\\
$(\alpha^1)^{(\mathfrak{t})}(X_{1(\mathfrak{s}_1)}, \dots ,
X_{n(\mathfrak{s}_n)})$, with
$(\alpha^1)^{(\mathfrak{t})}(X_{1(\mathfrak{s}_1)}, \dots ,
X_{n(\mathfrak{s}_n)})\in \mathfrak{G}_{\mathfrak{t}}$ and
$X_{i(\mathfrak{s}_a)}\in \mathfrak{G}_{\mathfrak{s}_a}$,
 $a=1,\dots,n$ \\ are known.

\medskip
\noindent
{\bf Proposition}. If $\alpha^1$ is a one-cocycle of a semisimple
$n$-Lie algebra, then
\begin{equation*}
(\alpha^1)^{(\mathfrak{t})}(X_{1(\mathfrak{s}_1)}, \dots ,
X_{n(\mathfrak{s}_n)})=0
\end{equation*}
when there are at least three indices
among the simple ideals $\mathfrak{t},\mathfrak{s}_1, \dots ,
\mathfrak{s}_n$ that are different {\it i.e.}, when the above
expression involves components in at least three different ideals.

{\it Proof}:
Without loss of generality we can choose
$\mathfrak{s}_n \neq \mathfrak{t}$. Since the ideal
$\mathfrak{G}_{\mathfrak{s}_n}$ is simple, there exist
$Y_{1(\mathfrak{s}_n)}, \dots , Y_{n(\mathfrak{s}_n)} \in
\mathfrak{G}_{\mathfrak{s}_n}$ such that
\begin{equation}
   X_{n(\mathfrak{s}_n)} = [Y_{1(\mathfrak{s}_n)}, \dots ,
   Y_{n(\mathfrak{s}_n)}] \ .
 \label{deformation13}
 \end{equation}
Then, the projection on the simple ideal
$\mathfrak{G}_{\mathfrak{t}}$ of the cocycle condition
(\ref{ntacohomology3}) tells us that
\begin{eqnarray}
\label{deformation14}
    & &\quad\quad\quad(\alpha^1)^{(\mathfrak{t})}(X_{1(\mathfrak{s}_1)}, \dots ,
    X_{n(\mathfrak{s}_n)}) = \nonumber\\
    & & \quad\quad\quad\quad =(\alpha^1)^{(\mathfrak{t})}(
    X_{1(\mathfrak{s}_1)}, \dots ,
    X_{n-1(\mathfrak{s}_{n-1})}, [Y_{1(\mathfrak{s}_n)}, \dots ,
   Y_{n(\mathfrak{s}_n)}])\nonumber\\
   & &  = - [ X_{1(\mathfrak{s}_1)}, \dots ,
    X_{n-1(\mathfrak{s}_{n-1})}, (\alpha^1)^{(\mathfrak{t})}
    (Y_{1(\mathfrak{s}_n)}, \dots ,
   Y_{n(\mathfrak{s}_n)})] \nonumber\\
  & &   + \sum_{a=1}^n [Y_{1 (\mathfrak{s}_n)}, \dots ,
   Y_{a-1(\mathfrak{s}_n)}, (\alpha^1)^{(\mathfrak{t})}
   (X_{1(\mathfrak{s}_1)}, \dots ,
    X_{n-1(\mathfrak{s}_{n-1})},Y_{a(\mathfrak{s}_n)}),
    Y_{a+1(\mathfrak{s}_n)}, \dots , Y_{n(\mathfrak{s}_n)} ]
    \\& &   + \sum_{a=1}^n (\alpha^1)^{(\mathfrak{t})}
    (Y_{1(\mathfrak{s}_n)}, \dots ,
   Y_{a-1(\mathfrak{s}_n)}, [X_{1(\mathfrak{s}_1)}, \dots ,
    X_{n-1(\mathfrak{s}_{n-1})},Y_{a(\mathfrak{s}_n)}],
    Y_{a+1(\mathfrak{s}_n)}, \dots , Y_{n(\mathfrak{s}_n)} ) =0
    \nonumber \; .
\end{eqnarray}
The first and third term in the last equality in
(\ref{deformation14}) vanish because we are assuming that at least
three different ideals are involved and the only possibility for a
non-vanishing result is that $\mathfrak{t}=\mathfrak{s}_1=\dots =
\mathfrak{s}_{n-1}$ and $\mathfrak{s}_1=\dots = \mathfrak{s}_{n-1} =
\mathfrak{s}_n$ respectively. The second term vanishes because
$\mathfrak{s}_n \neq \mathfrak{t}$ $\square$

\medskip
The previous Proposition allows us to simplify
(\ref{deformation12}) for the one-cocycle $\alpha^1$ to the case
where one or two different ideals are involved. This means that
$\alpha^1(X_1,\dots X_n)$ gives rise to the following terms:
\begin{eqnarray}
   & & \alpha^1(X_1,\dots ,X_n) =  \sum^k_{\mathfrak{s} =1}
    (\alpha^1)^{(\mathfrak{s})}(X_{1(\mathfrak{s})}, \dots ,
    X_{n(\mathfrak{s})})\nonumber \\
    & &\quad + \sum^k_{\mathfrak{s}\neq \mathfrak{t}}
    (\alpha^1)^{(\mathfrak{t})}(X_{1(\mathfrak{s})}, \dots ,
    X_{n(\mathfrak{s})}) \nonumber\\
    & & \quad + \sum^k_{\mathfrak{s}\neq \mathfrak{t}} \sum_{a=1}^n
    (\alpha^1)^{(\mathfrak{s})}(X_{1(\mathfrak{s})}, \dots ,
    X_{a-1(\mathfrak{s})},
    X_{a(\mathfrak{t})},X_{a+1(\mathfrak{s})}, \dots ,
    X_{n(\mathfrak{s})})\ .
 \label{deformation15}
\end{eqnarray}

\medskip
\noindent
{\bf Proposition}. Let $X_{n(\mathfrak{t})}=[Y_{1(\mathfrak{t})},
\dots , Y_{n(\mathfrak{t})}]$. If $\mathfrak{t}\neq \mathfrak{s}$,
then
\begin{equation}
   (\alpha^1)^{(\mathfrak{s})}(X_{1(\mathfrak{s})}, \dots ,
    X_{n-1(\mathfrak{s})},X_{n(\mathfrak{t})} ) = -
    [X_{1(\mathfrak{s})}, \dots ,
    X_{n-1(\mathfrak{s})},(\alpha^1)^{(\mathfrak{s})}
    (Y_{1(\mathfrak{t})}, \dots , Y_{n(\mathfrak{t})})] \ .
 \label{deformation16}
 \end{equation}
\medskip

 {\it Proof}:
 First, we notice that all the terms in the last line of
 (\ref{deformation15}) have the structure of the $l.h.s.$ of
 (\ref{deformation16}), and therefore eq. (\ref{deformation16})
 will apply to all of them. To prove now this relation,
 we again make use of eq.~\eqref{deformation14},
 particularized to the case where the a priori different ideals
 labelled  $\mathfrak{t},\mathfrak{s}_1,\dots ,
 \mathfrak{s}_n$  are such that the first  $n$ are equal
to $\mathfrak{s}$, say, and the $(n+1)$-th one $\mathfrak{s}_n$
 is different, say $\mathfrak{t}$. Then,
\begin{eqnarray}
    & &(\alpha^1)^{(\mathfrak{s})}(X_{1(\mathfrak{s})}, \dots ,
    X_{n-1(\mathfrak{s})}, X_{n(\mathfrak{t})} ) =\nonumber\\
    & &= (\alpha^1)^{(\mathfrak{s})}(
    X_{1(\mathfrak{s})}, \dots ,
    X_{n-1(\mathfrak{s})}, [Y_{1(\mathfrak{t})}, \dots ,
   Y_{n(\mathfrak{t})}])\nonumber\\
   = &-& [ X_{1(\mathfrak{s})}, \dots ,
    X_{n-1(\mathfrak{s})}, (\alpha^1)^{(\mathfrak{s})}
    (Y_{1(\mathfrak{t})}, \dots ,
   Y_{n(\mathfrak{t})})] \nonumber\\
   &+& \sum_{a=1}^n [Y_{1(\mathfrak{t})}, \dots ,
   Y_{a-1(\mathfrak{t})}, (\alpha^1)^{(\mathfrak{s})}
   (X_{1(\mathfrak{s})}, \dots ,
    X_{n-1(\mathfrak{s})},Y_{a(\mathfrak{t})}),
    Y_{a+1(\mathfrak{t})}, \dots , Y_{n(\mathfrak{t})} ]
    \nonumber\\
     & + &\sum_{a=1}^n (\alpha^1)^{(\mathfrak{s})}
    (Y_{1(\mathfrak{t})}, \dots ,
   Y_{a-1(\mathfrak{t})}, [X_{1(\mathfrak{s})}, \dots ,
    X_{n-1(\mathfrak{s})},Y_{a(\mathfrak{t})}],
    Y_{a+1(\mathfrak{t})}, \dots , Y_{n(\mathfrak{t})} ) \ ,
\label{deformation17}
\end{eqnarray}
and we see that the terms in the last two lines vanish, which proves
eq. (\ref{deformation16}) $\square$

\medskip

Using the two previous propositions, we now prove the
following

\medskip
\noindent {\bf Theorem}. Let $(\alpha^1)$ be a one-cocycle for the deformation
cohomology of a semisimple $n$-Lie algebra $\mathfrak{G}$, eq.
(\ref{adcoho}). Then there exists a zero-cochain $\beta$ such that
\begin{eqnarray}
  \alpha^1(X_1,\dots ,X_n)  &=& (\delta \beta)(X_1,\dots ,X_n)
    \nonumber\\
    &=& - \beta([X_1,\dots ,X_n]) +  \sum_{a=1}^n [X_1,\dots ,X_{a-1},
    \beta(X_a),X_{a+1}, \dots X_n] \; .
 \label{deformation18}
 \end{eqnarray}
Therefore, any semisimple FA is rigid.
\medskip

 {\it Proof}:
 Let us first consider the
 $\mathfrak{G}_{\mathfrak{s}}$-valued one-cochain with arguments
 in $\mathfrak{G}_{\mathfrak{s}}$ given by $
 (\alpha^1)^{(\mathfrak{s})}(X_{1(\mathfrak{s})},\dots
 ,X_{n(\mathfrak{s})})$ which corresponds to the first line in
 (\ref{deformation15}). If $\alpha^1$ is a cocycle, then
 so is $(\alpha^1)^{(\mathfrak{s})}$. But since
 $\mathfrak{G}_{\mathfrak{s}}$ is simple there exists a
 $\mathfrak{G}_{\mathfrak{s}}$-valued zero-cochain that takes
 arguments in $\mathfrak{G}_{\mathfrak{s}}$,
 ${\beta^{(\mathfrak{s})}}_{(\mathfrak{s})}$, such that eq.
 (\ref{ntacohomology6})
 reads
 \begin{eqnarray}
   & &(\alpha^1)^{(\mathfrak{s})}(X_{1(\mathfrak{s})},\dots
   ,X_{n(\mathfrak{s})})  =  -
   {\beta^{(\mathfrak{s})}}_{(\mathfrak{s})}([X_{1(\mathfrak{s})},
   \dots ,X_{n(\mathfrak{s})}])
    \nonumber\\
    & & \quad\quad+  \sum_{a=1}^n [X_{1(\mathfrak{s})},\dots ,
    X_{{a-1(\mathfrak{s})}},
    {\beta^{(\mathfrak{s})}}_{(\mathfrak{s})}(X_{a(\mathfrak{s})}),
    X_{a+1(\mathfrak{s})}, \dots X_{n(\mathfrak{s})}]\ .
 \label{deformation19}
 \end{eqnarray}
 Consider now the
 $\mathfrak{G}_{\mathfrak{t}}$-valued one-cochain with arguments
 in $\mathfrak{G}_{\mathfrak{s}}$ ($\mathfrak{t} \neq
 \mathfrak{s}$) given by $
 (\alpha^1)^{(\mathfrak{t})}(X_{1(\mathfrak{s})},\dots
 ,X_{n(\mathfrak{s})})$ (eq. (\ref{deformation15}), second line). This
 part also satisfies the one-cocycle condition since $\alpha^1$ is a
 cocycle, but the one cocycle condition (\ref{ntacohomology3}) reduces to the terms
 \begin{eqnarray}
    & & (\alpha^1)^{(\mathfrak{t})}(X_{1(\mathfrak{s})} ,\dots , X_{n-1(\mathfrak{s})},
    [Y_{1(\mathfrak{s})},\dots , Y_{n(\mathfrak{s})}])\nonumber\\
    & & \quad=
    \sum_{a=1}^n (\alpha^1)^{(\mathfrak{t})}(Y_{1(\mathfrak{s})} ,\dots , Y_{a-1(\mathfrak{s})},
    [X_{1(\mathfrak{s})} ,\dots , X_{n-1(\mathfrak{s})},
    Y_{a(\mathfrak{s})}], Y_{a+1(\mathfrak{s})}, \dots , Y_{n(\mathfrak{s})})\
    .
\label{deformation19a}
\end{eqnarray}
This is the one-cocycle condition that we considered already for the
trivial action (eq. (\ref{cohomology3})), and we know that then
$\alpha^1$ may be generated by a zero-cochain $\beta$, namely
 \begin{equation}
  (\alpha^1)^{(\mathfrak{t})}(X_{1(\mathfrak{s})},\dots
 ,X_{n(\mathfrak{s})})=(\delta{\beta^{(\mathfrak{t})}}_{(\mathfrak{s})})
 (X_{1(\mathfrak{s})},\dots
 ,X_{n(\mathfrak{s})})
  = - {\beta^{(\mathfrak{t})}}_{(\mathfrak{s})}
 ([X_{1(\mathfrak{s})},
   \dots ,X_{n(\mathfrak{s})}]) \ ,
 \label{deformation20}
\end{equation}
where the $\mathfrak{G}_{\mathfrak{t}}$-valued zero-cochain takes
arguments in $\mathfrak{G}_{\mathfrak{s}}$, hence the notation
${\beta^{(\mathfrak{t})}}_{(\mathfrak{s})}$.

Using these zero-cochains, we check that $\alpha^1=\delta \beta$
with
\begin{equation}
    \beta(X) := \sum_{\mathfrak{s},\mathfrak{t}=1}^{k}
    {\beta^{(\mathfrak{t})}}_{(\mathfrak{s})} (X_{(\mathfrak{s})})\ .
   \label{deformation21}
\end{equation}
Indeed, we have, from the definition of $\delta\beta$ (eq.
(\ref{ntacohomology6})),
\begin{eqnarray}
  & & \delta \beta (X_{1},\dots
   ,X_{n})  =  -
   \beta([X_{1}, \dots ,X_{n}])
    \nonumber\\
    & & \quad\quad+  \sum_{a=1}^n [X_{1},\dots ,
    X_{a-1},
    \beta(X_{a}),
    X_{a+1}, \dots X_{n}]
    \nonumber\\
    & & \quad= - \sum_{\mathfrak{s},\mathfrak{t}=1}^{k}
    {\beta^{(\mathfrak{t})}}_{(\mathfrak{s})}([X_{1},
    \dots ,X_{n}]_{(\mathfrak{s})}) \nonumber\\
    & & \quad\quad+ \sum_{a=1}^n \sum_{\mathfrak{s},\mathfrak{t}=1}^{k}
    [X_{1},\dots ,X_{a-1},
    {\beta^{(\mathfrak{t})}}_{(\mathfrak{s})}(X_{a(\mathfrak{s})}),
    X_{a+1}, \dots X_{n}] \nonumber \\
 & & \quad=  - \sum_{\mathfrak{s},\mathfrak{t}=1}^{k}
    {\beta^{(\mathfrak{t})}}_{(\mathfrak{s})}([X_{1(\mathfrak{s})},
    \dots ,X_{n(\mathfrak{s})}]) \nonumber\\
    & &\quad\quad + \sum_{a=1}^n \sum_{\mathfrak{s},\mathfrak{t}=1}^{k}
    [X_{1(\mathfrak{t})},\dots ,X_{a-1(\mathfrak{t})},
    {\beta^{(\mathfrak{t})}}_{(\mathfrak{s})}(X_{a(\mathfrak{s})}),
    X_{a+1(\mathfrak{t})}, \dots X_{n(\mathfrak{t})}]
    \ .
 \label{deformation22}
 \end{eqnarray}

The proof is now completed by checking that the coboundary
(\ref{deformation21}) generates the one-cocycle $\alpha^1$ in
(\ref{deformation15}) since, as argued there, all other cases give
zero. This is achieved by means of the Proposition
expressed by eq.~\eqref{deformation16}.
Using eq.~\eqref{deformation16} in the last terms of (\ref{deformation15}),
eq. (\ref{deformation19}) in the first term of (\ref{deformation15})
and eq. (\ref{deformation20}) both in the second term plus in the
terms which have now appeared from the $r.h.s$ of
(\ref{deformation16}) we obtain that the one-cocycle $\alpha^1$
gives rise to
\begin{eqnarray}
   & & \alpha^1 (X_{1},\dots
   ,X_{n})  =  - \sum_{\mathfrak{s}=1}^k
   {\beta^{(\mathfrak{s})}}_{(\mathfrak{s})}
([X_{1(\mathfrak{s})}, \dots ,X_{n(\mathfrak{s})}])
    \nonumber\\
    & &\quad\quad\quad\quad +  \sum_{\mathfrak{s}=1}^k \sum_{a=1}^n
    [X_{1(\mathfrak{s})},\dots ,
X_{a-1(\mathfrak{s})},
{\beta^{(\mathfrak{s})}}_{(\mathfrak{s})}(X_{a(\mathfrak{s})}),
    X_{a+1(\mathfrak{s})}, \dots X_{n(\mathfrak{s})}]
    \nonumber\\
    & &\quad\quad\quad\quad - \sum_{\mathfrak{s}\neq \mathfrak{t}}^{k}
    {\beta^{(\mathfrak{t})}}_{(\mathfrak{s})}([X_{1(\mathfrak{s})},
    \dots ,X_{n(\mathfrak{s})}]) \nonumber\\
    & &\quad\quad\quad\quad + \sum_{\mathfrak{s}\neq \mathfrak{t}}^{k} \sum_{a=1}^n
    [X_{1(\mathfrak{s})},\dots ,X_{a-1(\mathfrak{s})},
    {\beta^{(\mathfrak{s})}}_{(\mathfrak{t})}([Y_{a1(\mathfrak{t})}, \dots ,
    Y_{an(\mathfrak{t})}]),
    X_{a+1(\mathfrak{s})}, \dots X_{n(\mathfrak{s})}] \nonumber \\
  & & \quad\quad\quad = -\sum_{\mathfrak{s}, \mathfrak{t}=1}^{k}
    {\beta^{(\mathfrak{t})}}_{(\mathfrak{s})}([X_{1(\mathfrak{s})},
    \dots ,X_{n(\mathfrak{s})}]) \nonumber\\
    & & \quad\quad\quad\quad + \sum_{\mathfrak{s}, \mathfrak{t}=1}^{k} \sum_{a=1}^n
    [X_{1(\mathfrak{s})},\dots ,X_{a-1(\mathfrak{s})},
    {\beta^{(\mathfrak{s})}}_{(\mathfrak{t})}(X_{a(\mathfrak{t})}),
    X_{a+1(\mathfrak{s})}, \dots X_{n(\mathfrak{s})}] \ ,
 \label{deformation23}
 \end{eqnarray}
since, again, $[Y_{a1(\mathfrak{t})}, \dots ,Y_{an(\mathfrak{t})} ]=
X_{a(\mathfrak{t})}$, which allows us to join the two double sums in
the second and fourth lines above into a single one. This reproduces
eq.~\eqref{deformation22} so that $\alpha^1=\delta \beta$, which
completes the proof $\square$

\section{An observation on $n$-Leibniz algebras, FAs, and cohomology}.

Consider the case of ordinary or $n=2$ {\it Leibniz algebras}
$\mathscr{L}$ \cite{Lod:93,Lod-Pir:93,Cuv:94, Lod-Pir:96}. These
algebras share with the Lie algebras a form of the JI identity, but not the
anticommutativity of the two-bracket. As a result, their characteristic
identity is the {\it Leibniz identity}, which retains the aspect
(b) of the JI for Lie algebras mentioned in the Introduction.
Specifically, a {\it Leibniz algebra} is a vector space
$\mathscr{L}$ endowed with a bilinear operation $\mathscr{L}\times
\mathscr{L}\rightarrow \mathscr{L}$ that satisfies
the {\it Leibniz identity}
\begin{equation}
\label{left-Lei-alg} [X,[Y,Z]]=[[X,Y],Z]+[Y,[[X,Z]]   \qquad \forall
X,Y,Z \in \mathscr{L} \quad ,
\end{equation}
Actually, this defines a {\it left} Leibniz algebra. There is a right
counterpart when the equation above is modified to correspond to
a right derivation: the right Leibniz identity reads
$[[X,Y],Z]=[[X,Z],Y]+[X,[Y,Z]]$ and accordingly defines a right
Leibniz algebra. Obviously, when the bracket is
anticommutative, the Leibniz algebra $\mathscr{L}$ becomes a Lie
algebra $\mathfrak{g}$ and both the left and right Leibniz identities
become one and the same Lie algebra JI.

 Similarly, $n$-Leibniz algebras $\mathfrak{L}$
\cite{Da-Tak:97,Cas-Lod-Pir:02} (see also \cite{Rot:05}) share with
the FAs the derivation property expressed by the $n$-Leibniz
identity, which follows the pattern of the FI of eq.~\eqref{intro2},
where now the brackets are $n$-Leibniz brackets and thus not fully
antisymmetric\footnote{This is the case of the three algebras
used in \cite{Che-Sae:08} in the context of the BLG model.}.
Thus, their characteristic identity has still
the structure of eq.~\eqref{Lie-setb} which, to stress the similarity with
the left character of the Leibniz identity above we
shall write in the form
\begin{equation}
\label{FI-n-Leib}
\mathscr{X}\cdot (\mathscr{Y}\cdot \mathscr{Z})  =
(\mathscr{X}\cdot\mathscr{Y}) \cdot \mathscr{Z}
 + \mathscr{Y}\cdot(\mathscr{X} \cdot \mathscr{Z})
 \qquad \forall
\mathscr{X}, \mathscr{Y}, \mathscr{Z} \in \otimes^{n-1}\mathfrak{L}
\; ,
\end{equation}
where the $n$-bracket involved in the definition of the composition
of fundamental objects (eq.~\eqref{intro4}) is now the $n$-Leibniz
bracket in $\mathfrak{L}$.  It is clear that the above
defines a (left) $n$-Leibniz algebra. FAs
$\mathfrak{G}$ may be viewed as as a particular case of
$n$-Leibniz algebras $\mathfrak{L}$, namely those with a
fully antisymmetric $n$-bracket and fundamental objects
in $\wedge^{n-1}\mathfrak{G}$ rather than in $\otimes^{n-1}\mathfrak{L}$.

   When we were discussing the various FA cohomology complexes, the
essential ingredients in their definition were the FI and the specific
action of $\mathfrak{G}$ on the cochains; in particular, the
skewsymmetry of the FA $n$-bracket was not needed for the nilpotency
of $\delta$ in, say, eq.~\eqref{n-cobop}. As a result, this $\delta$
also defines a coboundary operator for the cohomology of a
$n$-Leibniz algebra $\mathfrak{L}$, in which the $p$-cochains are
now elements $\alpha\in (\otimes^{(n-1)}\mathfrak{L}^*)\otimes
\mathop{\cdots}\limits^p\otimes
(\otimes^{(n-1)}\mathfrak{L}^*)\otimes \mathfrak{L}^*=
\otimes^{p(n-1)+1}\mathfrak{L}^*.$
The key ingredient that guarantees the nilpotency of
the coboundary operator is still the (left) identity that
follows from eq.~\eqref{intro2}, which implies
eqs.~\eqref{Lie-setb}, \eqref{Lie-n-Lie}, etc.  The essential difference
between the $n$-Lie algebra and $n$-Leibniz algebra cohomology
complexes for the trivial representation, unimportant for the
nilpotency of $\delta$, is that
$\mathscr{X}\in \wedge^{n-1}\mathfrak{G}$ for a $n$-Lie algebra
and $\mathscr{X}\in \otimes^{n-1}\mathfrak{L}$ for a Leibniz algebra,
and the fact the brackets that appear in the composition of
fundamental objects in eq.~\eqref{intro4} are, respectively,
$n$-Lie and $n$-Leibniz brackets.

   Thus, with the appropriate changes in the definition of the
$p$-cochain spaces $C^p$, the coboundary operator in
eq.~\eqref{n-cobop} defines the corresponding cohomologies for
$n$-Lie $\mathfrak{G}$ and $n$-Leibniz $\mathfrak{L}$ algebras
adapted to the central extension problem, which corresponds to the
trivial action.
 Similar considerations apply to the $n$-Leibniz algebra
cohomology adapted to the deformation problem already discussed for
the FAs (eqs.~\eqref{adcoho}, \eqref{p-coc-ftal}), but we shall
not consider this further here. For an ordinary Leibniz algebra
$\mathcal{L}$, for instance, generalizing to the
case where the action is given through a representation
\cite{Lod:93,Lod-Pir:96} $\rho$ on $\mathscr{A}$ and reverting
(since $n=2$) to the notation where $p$ indicates the number of
algebra elements on which $\alpha^p$ takes arguments, $\alpha^p\in
C^p(\mathscr{L},\mathscr{A})=\hbox{Hom}(\otimes^p
\mathscr{L},\mathscr{A})$ , eq.~\eqref{adcoho} leads to
\begin{equation}
\label{deform-defb-n2}
\begin{aligned}
(\delta\alpha^p) &
(X_1,\dots,X_p,X_{p+1})= \\
&\sum_{1\leq j<k}^{p+1} (-1)^j
\alpha^p(X_1,\dots,\widehat{X_j},\dots,X_{k-1},
[X_j, X_k], X_{k+1},\dots,X_{p+1})\\
 + &\sum_{j=1}^{p} (-1)^{j+1} \rho(X_j) \cdot
 \alpha^p(X_1,\dots,\widehat{X_j}\dots,X_{p+1}) \\
 +&(-1)^{p+1}
 \alpha^p(X_1,\dots, X_p) \cdot\rho( X_{p+1}) \quad ,
\end{aligned}
\end{equation}
which coincides with the coboundary operator for the Leibniz algebra
cohomology complex $(C^\bullet(\mathscr{L},\mathscr{A}), \delta)$
\cite{Lod-Pir:93,Lod:93,Cas-Lod-Pir:02} (there given for
right Lebiniz algebras). We note that if $\rho$ is a symmetric
representation \cite{Lod:93,Lod-Pir:96}, the $\rho(X_{p+1})$ in the
last term above may be moved to the left and the resulting contribution
may then be added as one more term to the third line by
enlarging the range of the sum.  The resulting expression
has then the same form of the expression that gives the action
of the Lie algebra cohomology coboundary operator on
$p$-cochains valued on a $\rho(\mathfrak{g})$-module
(see, {\it e.g.} \cite{CUP}).

Thus, the FA cohomologies defined by the coboundary operators
\eqref{n-cobop} that define the FA cohomology complexes (and the
corresponding homology) constitute simply the translation of the
$n$-Leibniz algebra $\mathfrak{L}$ cohomology complexes to the
$n$-Lie algebra $\mathfrak{G}$ case. In fact, the previous discussion
shows that since Leibniz algebras
largely underly the structural cohomological properties of the FAs,
the FA cohomology complexes could also have been found from those
for the $n$-Leibniz algebras by demanding full skewsymmetry for the
$n$-Leibniz bracket to become the $n$-bracket of a FA, and by
modifying accordingly the definition of the cochains to account for
the skewsymmetry of the fundamental objects of a FA.
\medskip

We conclude this section with a remark.
Let $\mathfrak{L}$ be a $n$-Leibniz algebra. The composition
$\mathscr{X}\cdot\mathscr{Y}$ of fundamental objects, rewritten as
$[\mathscr{X},\mathscr{Y}]$, has the properties of a
(non-antisymmetric) Leibniz algebra commutator. Indeed,
eq.~\eqref{Lie-setb} also holds for $\mathfrak{L}$, and with
the notation
$\mathscr{X}\cdot\mathscr{Y}\equiv[\mathscr{X}\,,\,\mathscr{Y}]$ it
takes the form
\begin{equation}
\label{FA-Leib} [\mathscr{X}\,,\,[\mathscr{Y}\,,\, \mathscr{Z}]]
=[[\mathscr{X}\,,\,\mathscr{Y}]\,,\,\mathscr{Z}]+ [\mathscr{Y}\,,\,
[\mathscr{X}\,,\,\mathscr{Z}]]\quad, \quad
\mathscr{X}\,,\,\mathscr{Y}\,,\, \mathscr{Z} \in
\otimes^{n-1}\mathfrak{L} \quad,
\end{equation}
where $[\mathscr{X}\,,\,\mathscr{Y}]\not=
-[\mathscr{Y}\,,\,\mathscr{X}]\,$ is a non-antisymmetric
two-bracket. Comparing with eq.~\eqref{left-Lei-alg}, we see that
\eqref{FA-Leib} defines an ordinary (left) Leibniz algebra
where the two entries in $[\;,\;]$ are
fundamental objects $\mathscr{X}\in \otimes^{n-1}\mathfrak{L}$.
Hence, given a
$n$-Leibniz algebra $\mathfrak{L}$, the linear space of the
fundamental objects endowed with the dot operation \eqref{intro4}
which defines a non-antisymmetric two-bracket
$[\mathscr{X},\mathscr{Y}]\equiv \mathscr{X}\cdot\mathscr{Y}$,
becomes a (here left) Leibniz algebra \cite{Da-Tak:97},
the {\it ordinary Leibniz algebra $\mathscr{L}$ associated with a
$n$-Leibniz algebra} $\mathfrak{L}$.

When the fundamental objects are those of a FA, $\mathscr{X},
\mathscr{Y} \in \wedge^{n-1}\mathfrak{G}$, and the $n$-bracket
involved in the definition of $\mathscr{X}\cdot\mathscr{Y}$ is
therefore the fully antisymmetric bracket
of a $n$-Lie algebra, the resulting $\mathscr{L}$ is
the {\it Leibniz algebra associated with the Filippov algebra
$\mathfrak{G}$}. For $n$=2, the Leibniz algebra $\mathscr{L}$
associated with the $n=2$ FA $\mathfrak{G}$ is in fact an
ordinary Lie algebra $\mathfrak{g}$ since the FA bracket is
skewsymmetric and the FA itself is an ordinary Lie algebra.
\medskip

\section{Final comments}

We have proved the analogue of Whitehead's lemma for $n$-Lie
algebras: semisimple Filippov algebras cannot be centrally extended
in a non-trivial way and furthermore they are rigid because
$H^1(\mathfrak{G},\mathfrak{G})=0$. Actually, Witehead's
lemma for ordinary Lie algebras is more general since it states that
the Lie algebra cohomology groups for a non-trivial action ($\rho\not=0$)
are trivial, $H_\rho^p(\mathfrak{g},V)=0\; \forall p\geq 0$,
when $\mathfrak{g}$ is semisimple. The analogous proof for
Filippov algebras $\mathfrak{G}$, using our procedure, would require
proving first the triviality of the corresponding higher order
cohomology groups for simple $n$-Lie algebras. This is much more involved,
but the fact that the analogue \cite{Kas:95a} of the Cartan-Killing form for
a simple $\mathfrak{G}$ is non-degenerate when considered as a
bilinear form on $\wedge^{n-1}\mathfrak{G}$, could help in the
analysis of all these higher order cohomology groups, which we
conjecture to be also trivial.

Another extension of our results would be to consider the
case of Leibniz's $n$-algebras $\mathfrak{L}$ mentioned in Sec.~6,
since their cohomological structure, being based on the $n$-Leibniz
identity, is similar. However, since the proof of the triviality of the relevant
FA cohomology groups relies heavily on the skewsymmetry of the
structure constants of the simple FAs, one would not expect
{\it e.g.} their rigidity property to extend automatically to the corresponding
Leibniz algebra case. In fact, it has been
shown\cite{F-O'F:08} that the simple euclidean three-Lie algebra $A_4$
may be infinitesimally deformed as a three-Leibniz algebra of
a specific type. This is not surprising at the light of the above
discussion, and indeed this increase of possibilities for deformations
has been observed already for ordinary Lie algebras when they are
treated as Leibniz algebras \cite{Fia-Man:08}; in other words, the
Leibniz algebra deformations of a Lie algebra are richer.
However, the mentioned Leibniz deformability of the simple
$A_4$ euclidean algebra may not extend to the other simple FAs
(the case $n$=3 is special), which may be rigid
under this type of $n$-Leibniz algebra deformations \cite{Az-Iz:09}.

\subsection*{Acknowledgments}

This work has been partially supported by research grants from the
Spanish Ministry of Science and Innovation (FIS2008-01980,
FIS2005-03989), the Junta de Castilla y
Le\'on (VA013C05) and EU FEDER funds.  

\vskip 1cm


\begin{thebibliography}{99}

\bibitem{AzPePB:96b}
J. A. de Azc\'arraga, A. M. Perelomov and J. C. P\'erez Bueno,
{\it The Schouten-Nijenhuis bracket, cohomology and generalized Poisson  structures},
J. Phys. {\bf A29}, 7993-8010 (1996) {\tt [arXiv:hep-th/9605067]}.

\bibitem{AzBu:96}
J.~A. de~Azc\'arraga and J.~C. P\'erez-Bueno, {\it Higher-order
simple Lie algebras}, Commun. Math. Phys. {\bf 184},  669-681 (1997)
{\tt [arXiv:hep-th/9605213]}.

\bibitem{RACSAM:98}
  J.~A.~de Azc\'arraga, J.~M.~Izquierdo and J.~C.~P\'erez Bueno,
{\it An introduction to some novel applications of Lie algebra cohomology in
mathematics and physics},
Rev. Real Acad. Cien. Exactas F\'{\i}s. Nat., Ser. A Mat., {\bf 95}, 225-248 (2001)
{\tt [arXiv:physics/9803046]}.

\bibitem{Han-Wac:95}
P. Hanlon and M. Wachs, {\it On Lie $k$-algebras}, Adv. in Math.
{\bf 113}, 206-236 (1995)

\bibitem{JLL:95}
J.-L. Loday, {\it La renaissance of op\'erades}, Sem. Bourbaki
{\bf 792}, 47-74 (1994-95).

\bibitem{Gne:95}
V. Gnedbaye, {\it Les alg\`ebres $k$-aires el leurs op\'erads}, C.
R. Acad. Sci. Paris, S\'erie I {\bf 321}, 147-152 (1995)

\bibitem{Mic-Vin:96}
P.W. Michor and A. M. Vinogradov, {\it $n$-ary Lie and associative
algebras}, Rend. Sem. Mat. Univ. Pol. Torino {\bf 53}, 373-392
(1996) {\tt [arXiv:math.QA/9801087]}.

\bibitem{Gne:97}
V. Gnedbaye, {\it Operads of $k$-ary algebras}, Contemp. Math. {\bf
202},83-114 (1997)

\bibitem{Lad.Sta:93}
T. Lada and J. Stasheff, {\it Introduction to SH Lie algebras
for physicists}, Int. J. Theor. Phys. {\bf 32}, 1087-1104 (1993)
{\tt [arXiv:hep-th/9209099]}.

\bibitem{Lad.Mar:95}
T.~Lada and M.~Markl,
{\it Strongly homotopy Lie algebras},
Commun. in Alg. {\bf 23}, 2147-2161 (1995).

\bibitem{Sta:97}
J.~Stasheff,
{\it Deformation theory and the Batalin-Vilkovisky master equation},
in {\it Deformation theory and symplectic geometry},
D. Sternheimer {\it et al.} eds., Kluwer Acad. Publ.,
p. 271-284, 1997
{\tt [arXiv:q-alg/9702012]}.

\bibitem{Be-La:09}
K. Bering and T. Lada,
{\it Examples of homotopy Lie algebras},
{\tt arXiv:0903.5433v1 [math.QA]}

\bibitem{Filippov}
V.~Filippov, {\it $n$-Lie algebras}, Sibirsk. Mat. Zh. {\bf 26},
126-140 (1985) (English transl.: Siberian Math. J.
  \textbf{26}, 879-891 (1985)).

\bibitem{Kas:87}
S.~M. Kasymov, {\it Theory of {$n$}-lie algebras}, Algebra i Logika
{\bf 26}, 277-297 (1987) (English transl.: Algebra and Logic
\textbf{26}, 155-166 (1988)).

\bibitem{Kas:95a}
S.~M. Kasymov, {\it Analog of the Cartan criteria for {$n$}-lie
algebras}, Algebra i Logika {\bf 34}, 274-287 (1995) (English
transl.: Algebra and Logic {\bf 34}, 147-154 (1995)).

\bibitem{Ling:93}
W.~X. Ling, {\it On the structure of $n$-{Lie} algebras}, PhD
thesis, Siegen, 1993.

\bibitem{Ba-La:06}
J. Bagger and N. Lambert, {\it Modelling multiple M2's}, {\em
Phys. Rev.} {\bf D75}, 045020 (2007)
{\tt arXiv:hep-th/0611108}.

\bibitem{Ba-La:07}
J.~Bagger and N.~Lambert,
{\it Gauge Symmetry and Supersymmetry of Multiple M2-Branes},
Phys. Rev. {\bf D77}, 065008 (2008)
{\tt [arXiv:0711.0955 [hep-th]]}.

\bibitem{Gustav:08}
A.~Gustavsson, {\it One-loop corrections to Bagger-Lambert
theory}, Nucl. Phys. {\bf B807}, 315-333 (2009)
{\tt arXiv:0805.4443 [hep-th]}.

\bibitem{Raam:08}
M. Van Raamsdonk, {\it Comments on the Bagger-Lambert theory and
multiple M2-branes}, JHEP {\bf 05}, 105 (2008)
{\tt arXiv:0803.3803 [hep-th]}.


\bibitem{Ba-La:08}
  J.~Bagger and N.~Lambert,
{\it Three-Algebras and N=6 Chern-Simons Gauge Theories},
Phys. Rev. {\bf D79}, 025002 (2009)
{\tt [arXiv:0807.0163 [hep-th]]}.


\bibitem{Che-Sae:08}
S.~A.~Cherkis and C.~S\"amann,
{\it Multiple M2-branes and Generalized 3-Lie algebras},
Phys. Rev. {\bf D78}, 066019 (2008) {\tt [arXiv:0807.0808 [hep-th]]}.

\bibitem{Go-Mi-Ru:08}
J.~Gomis, G.~Milanesi and J.~G.~Russo,
{\it Bagger-Lambert Theory for General Lie Algebras},
JHEP {\bf 0806}, 075 (2008) {\tt [arXiv:0805.1012 [hep-th]]}.

\bibitem{Ga-Gu:08}
J.~P.~Gauntlett and J.~B.~Gutowski,
{\it Constraining maximally supersymmetric membrane actions},
{\tt  arXiv:0804.3078 [hep-th]}.

\bibitem{Pap:08}
  G.~Papadopoulos,
{\it M2-branes, 3-Lie algebras and Pl\"ucker relations},
  JHEP {\bf 0805}, 054 (2008)
{\tt [arXiv:0804.2662 [hep-th]]}.

\bibitem{deMe-Fi-M-E-Rit:09}
  P.~de Medeiros, J.~Figueroa-O'Farrill, E.~M\'endez-Escobar and P.~Ritter,
{\it Metric 3-Lie algebras for unitary Bagger-Lambert theories},
  JHEP {\bf 0904}, 037 (2009)
{\tt [arXiv:0902.4674 [hep-th]]}.

\bibitem{A-B-J-M:08}
O.~Aharony, O.~Bergman, D.~L.~Jafferis and J.~Maldacena,
{\it N=6 superconformal Chern-Simons-matter theories,
M2-branes and their gravity duals},
JHEP {\bf 0810}, 091 (2008)
{\tt [arXiv:0806.1218 [hep-th]]}.

\bibitem{Jac:49}
N. Jacobson,
{\it Lie and {J}ordan triple systems},
Amer. J. Math., {\bf 71}, 149-170, (1949)

\bibitem{Oku-Kam:96}
S.~Okubo and N.~Kamiya,
{\it Jordan-Lie superalgebra and Jordan-Lie triple system},
Univ. Rocherster preprint UR-1467 (1996)

\bibitem{Oku:03}
S.~Okubo,
{\it Construction of Lie superalgebras from triple product systems},
AIP Conf. Proc. {\bf 687},33 (2003).

\bibitem{Ker:08}
R. Kerner,
{\it Ternary and non-associative structures},
Int. J. of Geom. Meth. in Mod. Phys.
{\bf 5}, 1265-1294 (2008)

\bibitem{Gers:63}
M.~Gerstenhaber, {\it On the deformation of rings and algebras},
Annals Math. {\bf 79}, 59-103 (1964).

\bibitem{Nij-Rich:67}
A.~Nijenhuis and R.~W. Richardson~Jr., {\it Deformation of Lie
algebra structures}, J. Math. Mech. {\bf 171}, 89-105 (1967)

\bibitem{Jac:79}
N.~Jacobson, {\it Lie algebras}. Dover Pub., N.Y., 1979.

\bibitem{CUP}
J.~A. de~Azc\'arraga and J.~M. Izquierdo, {\it Lie groups, Lie
algebras,  cohomology and some applications in physics},
Cambridge University Press, Cambridge, UK, 1995.

\bibitem{Gau:96}
P.~Gautheron, {\it Some remarks concerning Nambu mehcanics},
Lett. Math.  Phys. {\bf 37}, 103-116 (1996).

\bibitem{Da-Tak:97}
Y.~L. Daletskii and L.~Takhtajan, {\it Leibniz and Lie algebra
structures for Nambu algebra}, Lett. Math. Phys. {\bf 39} 127-141 (1997).

\bibitem{Rot:05}
M.~Rotkiewicz, {\it Cohomology ring of $n$-Lie algebras},
Extracta Math. {\bf 20}, 219-232 (2005).

\bibitem{Tak:94}
L. Takhtajan, {\it A higher order analog of the Chevalley-Eilenberg complex
and the deformation theory of $n$-algebras},
St. Petersburg Math. J. {\bf 6}, (1994) (English transl. {\bf 6}, 429-437 (1995)

\bibitem{Nambu:73}
Y.~Nambu, {\it Generalized Hamiltonian dynamics}, Phys. Rev.
{\bf D7},  2405-2414 (1973).

\bibitem{Sa-Va:92}
D.~Sahoo and M.~C. Valsakumar,
{\it Nambu mechanics and its quantization},
Phys. Rev. {\bf A46}, 4410-4412 (1992)

\bibitem{Sa-Va:93}
D.~Sahoo and M.~C. Valsakumar, {\it Algebraic structure of Nambu
mechanics}, Pramana {\bf 40}, 1-16 (1993).

\bibitem{Tak:93}
L.~Takhtajan, {\it On Foundation of the generalized Nambu
mechanics}, Commun. Math. Phys. {\bf 160}, 295-316 (1994),
{\tt[arXiv:hep-th/9301111]}.

\bibitem{Fil:98}
V.~Filippov, {\it On {$n$}-Lie algebra of jacobians}, Sibirsk. Mat.
Zh. {\bf 39}, no.~3, 660-669, (1998) (English transl.: Siberian
Math. J. \textbf{39}, 573-581 (1998)).

\bibitem{Mu-Sud:76}
N. Mukunda and E. G. C. Sudarshan, {\it Relation between Nambu
and Hamiltonian mechanics}, Phys. Rev. {\bf D13}, 2846-2850 (1976).

\bibitem{Cha:95}
R. Chatterjee, {\it Dynamical symmetries and Nambu mechanics}",
Lett. Math. Phys. {\bf 36}, 117-126 (1996) {\tt [arXiv:hep-th/9501141]}.

\bibitem{Cha-Tak:95}
R. Chatterjee and L. Takhtajan, {\it Aspects of classical and
quantum Nambu mechanics}, Lett. Math. Phys.{\bf 37}, 475-482 (1996)
{\tt [arXiv:hep-th/9507125]}.

\bibitem{Ale.Guh:96}
D. Alekseevsky and P. Guha, {\it On decomposability of
Nambu-Poisson tensor}, Acta Math. Univ. Comenianae {\bf LXV}, 1-9
(1996)

\bibitem{Hie:97}
J. Hietarinta, {\it Nambu tensors and commuting vector fields}, J.
Phys.{\bf A30}, L27-L23 (1997)

\bibitem{Ma-Vi-Vi:97}
G.~Marmo, G.~Vilasi and A.~M.~Vinogradov,
{\it The local structure of n-Poisson and n-Jacobi manifolds},
J.\ Geom.\ Phys.\  {\bf 25}, 141 (1998)
{\tt [arXiv:physics/9709046]}.

\bibitem{Vai:99}
I. Vaisman, {\it A survey on Nambu-Poisson brackets}, Acta Math.
Univ. Comenianae {\bf LXVIII}, 213-241 (1999) {\tt [arXiv:math/9901047]}.

\bibitem{Mi-Vi:00}
P.W. Michor and  A. M. Vinogradov, {\it A note on $n$-ary Poisson
brackets}, Rend. Circ. Mat. di Palermo Ser. II, suppl. {\bf 63},
165-172 (2000) {\tt [arXiv:math.SG/9901117]}.

\bibitem{Cu-Za:02}
T. Curtright, C. K. Zachos, {\it Classical and quantum Nambu
mechanics}, Phys. Rev. {\bf D68}, 085001-1-29 (2003)
{\tt [arXiv:hep-th/0212267]}.

\bibitem{Be-Se-Ta-To:90}
  E.~Bergshoeff, E.~Sezgin, Y.~Tanii and P.~K.~Townsend,
{\it Super $p$-Branes as gauge theories of volume preserving diffeomorphisms},
Annals Phys. {\bf 199}, 340-365 (1990).

\bibitem{Ho:96}
J.~Hoppe, {\it On M-Algebras, the quantisation of Nambu-mechanics, and volume preserving
diffeomorphisms}, Helv. Phys. Acta {\bf 70}, 302-317 (1997)
{\tt [arXiv:hep-th/9602020]}.

\bibitem{Ho-Hou-Ma:08}
P.~M.~Ho, R.~C.~Hou and Y.~Matsuo,
{\it Lie 3-algebra and multiple M2-branes},
JHEP {\bf 0806}, 020 (2008)
{\tt [arXiv:0804.2110 [hep-th]]}.

\bibitem{Sochi:08}
C.~Sochichiu,
{\it On Nambu-Lie 3-algebra representations},
{\tt arXiv:0806.3520 [hep-th]}.

\bibitem{Ho-Ma:08}
P.~M.~Ho and Y.~Matsuo,
{\it M5 from M2}
JHEP {\bf 0806}, 105 (2008)
{[arXiv:0804.3629 [hep-th]]}.

\bibitem{Ba-To:08}
I.~A.~Bandos and P.~K.~Townsend,
{\it SDiff gauge theory and the M2 condensate},
JHEP {\bf 0902}, 013 (2009)
{\tt [arXiv:0808.1583 [hep-th]]}.

\bibitem{AzPePB:96a}
J. A. de Azc\'arraga, A. M.  Perelomov, and J. C. P\'erez Bueno,
{\it New generalized Poisson structures}, J. Phys. {\bf A29},
L151-L157 (1996) {\tt [arXiv:q-alg/9601007]}.

\bibitem{AIP-B:97}
J. A. de Azc\'arraga, J. M Izquierdo  and J. C. P\'erez Bueno,
{\it On the generalizations of Poisson structures},
J. Phys. {\bf A30}, L607-L616 (1997) {\tt [arXiv:hep-th/9703019]}.

\bibitem{Iba.Leo.Mar.Die:97}
R. Ib\'a{\~n}ez M. de Le\'on, J. C. Marrero and and D. Mart\'{\i}n
de Diego, {\it Dynamics of generalized Poisson and Nambu-Poisson brackets},
J. Math. Phys. {\bf 38}, 2332-2344 (1997)

\bibitem{Iba.Leo.Mar:97}
R. Ib\'a{\~n}ez M. de Le\'on, J. C. Marrero,
{\it Homology and cohomology on generalized Poisson manifolds},
J. Phys. {\bf A31}, 1253-1266 (1998)

\bibitem{A-I-Pe-PB:96}
J.~A.~de Azc\'arraga, J.~M.~Izquierdo, A.~M.~Perelomov and J.~C.~P\'erez Bueno,
{\it The $Z_2$-graded Schouten-Nijenhuis bracket and generalized super-Poisson
structures}, J. Math. Phys. {\bf 38}, 3735 (1997)
{\tt [arXiv:hep-th/9612186]}.

\bibitem{Stern:98}
D. Sternheimer, {\it Deformation quantization: twenty years after},
AIP Conf. Proc., {\bf 453}, 107-145 (1998) {\tt [arXiv:math/9809056]}.
%

\bibitem{Cu-Za:03b}
T. Curtright and C. K. Zachos, {\it Quantizing Dirac and Nambu
brackets}, AIP Conf. Proc. {\bf 672}, 165-182 (2003)
{\tt [arXiv:hep-th/0303088]}.

\bibitem{Cu-Fa-Ji-Me-Za:09}
  T.~Curtright, D.~Fairlie, X.~Jin, L.~Mezincescu and C.~Zachos,
{\it Classical and Quantal Ternary Algebras},
{\tt arXiv:0903.4889 [hep-th]}.

\bibitem{Lod:93}
J.-L. Loday, {\it Une version non-commutative des alg\`ebres de
Lie}, L'Ens. Math. {\bf 39}, 269-293 (1993).

\bibitem{Lod-Pir:93}
J. L. Loday and T. Pirashvili,
{\it Universal enveloping algebras of Leibniz algebras and (co)homology},
Mat. Annalen {\bf 296}, 139-158 (1993).

\bibitem{Cuv:94}
C. Cuvier, {\it Alg\`ebres de Leibnitz: definitions,
propri\'et\'es}, Ann. Scient. \'Ec. Norm. Sup. {\bf 27}, 1-45 (1994)

\bibitem{Lod-Pir:96}
J.-L. Loday and T. Pirashvili, {\it Leibniz representations of Lie
algebras}, J. Alg. {\bf 181}, 414-425 (1996)

\bibitem{Cas-Lod-Pir:02}
J. M. Casas, J.-L. Loday and T. Pirashvili,
{\it Leibniz $n$-algebras}, Forum Math. {\bf 14},189-207 (2002).


\bibitem{F-O'F:08}
  J.~M.~Figueroa-O'Farrill,
{\it Three lectures on 3-algebras},
{\tt arXiv:0812.2865 [hep-th]}.

\bibitem{Fia-Man:08}
A. Fialowski and A. Mandal, {\it Leibniz algebra deformations of a
Lie algebra}, J. Math. Phys. {\bf 49}, 093511-1-11 (2008)
{\tt[arXiv:0802.1263 [math.KT]]}

\bibitem{Az-Iz:09}
J. A. de Azc\'arraga and J. M. Izquierdo,
{\it On Leibniz deformations and rigidity of simple
$n$-Lie algebras}, in preparation.

\end{thebibliography}
\end{document}